\documentclass[12pt,a4paper]{article}
\usepackage{graphicx}
\usepackage{amsmath, amssymb}
\usepackage{natbib}
\usepackage{dsfont}
\usepackage{hyperref}
\usepackage{url}
\usepackage{booktabs,tabularx, colortbl}
\usepackage{authblk} % For author's and affiliation
\usepackage{color} 
\usepackage{enumitem}
\usepackage{arydshln}
\usepackage{cleveref}
\usepackage{multirow}
\usepackage{multicol}
\usepackage{setspace}
\usepackage{verbatim}
\def \dsP {\text{$\mathds{P}$}}

\DeclareMathOperator{\logit}{logit}

\DeclareMathOperator{\Cov}{Cov}
\DeclareMathOperator{\Cor}{Cor}
\DeclareMathOperator{\Var}{Var}

\def \xvec {\text{\boldmath$x$}}    \def \mX {\text{\boldmath$X$}}
\def \yvec {\text{\boldmath$y$}}    \def \mY {\text{\boldmath$Y$}}

\def \betavec         {\text{\boldmath$\beta$}}

\def \thetavec        {\text{\boldmath$\theta$}}

\def \muvec           {\text{\boldmath$\mu$}}

\def \mSigma   {\mathbf{\Sigma}}

\def \nullvec {\mathbf{0}}

\title{\centering Boosting Multivariate Structured Additive Distributional Regression Models}

\author[1]{Annika Strömer}
\author[2]{Nadja Klein}
\author[1]{Christian Staerk}
\author[1]{Hannah Klinkhammer}
\author[1]{Andreas Mayr}

\small{
\affil[1]{Department of Medical Biometrics, Informatics and Epidemiology, University Hospital Bonn, Germany}
\affil[2]{Emmy Noether Research Group in Statistics and Data Science, Humboldt-Universit\"at zu Berlin, Berlin, Germany}
}

\date{} % Leer damit kein Datum erscheint

\begin{document}

\def\spacingset#1{\renewcommand{\baselinestretch}%
{#1}\small\normalsize} \spacingset{1}
\setlength{\parindent}{0em} % Damit der nach Absatz nicht einrückt
 
\maketitle
\thispagestyle{empty}
\begin{abstract}
We develop a model-based boosting approach for multivariate distributional regression within the framework of generalized additive models for location, scale, and shape. Our approach enables the simultaneous modeling of all distribution parameters of an arbitrary parametric distribution of a multivariate response conditional on explanatory variables, while being applicable to potentially high-dimensional data. Moreover, the boosting algorithm incorporates data-driven variable selection, taking various different types of effects into account. As a special merit of our approach, it allows for modelling the association between multiple continuous or discrete outcomes through the relevant covariates. After a detailed simulation study investigating estimation and prediction performance, we demonstrate the full flexibility of our approach in three diverse biomedical applications. The first is based on high-dimensional genomic cohort data from the UK Biobank, considering a bivariate binary response (chronic ischemic heart disease and high cholesterol). Here, we are able to identify genetic variants that are informative for the association between cholesterol and heart disease. The second application considers the demand for health care in Australia with the number of consultations and the number of prescribed medications as a bivariate count response. The third application analyses two dimensions of childhood undernutrition in Nigeria as a bivariate response and we find that the correlation between the two undernutrition scores is considerably different depending on the child's age and the region the child lives in.

%A key finding is that \textbf{XXXX}.

%Our method shows promising results for modeling the relationships between these two phenotypes in terms of selecting and estimating effects of explanatory variables also for the correlation parameter.
\end{abstract}
 {\bf Keywords:} GAMLSS; multivariate Gaussian distribution; multivariate logit model; multivariate Poisson distribution; model-based boosting; semiparametric regression.

\newpage

\setcounter{page}{1}

\spacingset{1.5}
\section{Introduction}
\setlength{\abovedisplayskip}{0.2cm}
\setlength{\belowdisplayskip}{0.2cm}
Many modern regression models relate certain characteristics of a univariate response distribution to explanatory variables. Examples include generalized additive models~\citep[GAMs;][]{HastieTibshirani1990,Woo2017} and quantile regression models~\citep{Koe2005}, where with the former the conditional expectation and with the latter conditional quantiles of a univariate response distribution are modeled by an additive decomposition of different covariate effects. When multiple outcomes are supposed to be analyzed as response variables, one may fit such univariate regression models separately for each outcome. However, in practice, the components of a multivariate response are often not (conditionally) independent, so that separate models induce a loss of information and might lead to potentially misleading conclusions. 

A well-known approach for the analysis of multivariate responses, particularly common in the economics literature, is called seemingly unrelated regression~\citep{zellner62}. This classical approach is restricted to linear predictors and constant covariance matrices not depending on the covariates; however, extensions to semiparametric predictors for the marginal means exist \citep{LanAdeFahSte2003}. Beyond that, multiple discrete responses (e.g., count data) can be analysed using seemingly unrelated Poisson regression~\citep{King1989} and non-linear predictors \citep{Gallant975}. 
Similar to the approach of \citet{zellner62} these models are limited in their flexibility, and only the expected value of the response is linked to the covariates~\citep{fiebig2001}.
% Modeling beyond the mean
A more flexible framework is provided by generalized additive models for location, scale and shape~\citep[GAMLSS;][]{gamlss}, in which each parameter of the conditional distribution is modeled by an additive predictor. The use of additive predictors for all distribution parameters, such as location, scale or skewness parameters allows to incorporate different effect types for the covariates in a very flexible way.

In high-dimensional data situations in which the number of predictors exceeds the number of observations ($p > n$), several classical estimation approaches are no longer feasible. Bayesian variable selection (e.g.,~\citealt{Zhu2012}) and penalized regression methods (e.g.,~\citealt{Wu2014,Liu2020}) have been proposed for multivariate modelling in high-dimensional data situations. 
Nevertheless, GAMLSS based on penalized likelihood estimation is currently only available for univariate response variables.
In contrast, \cite{KleKneKlaLan2015} extended this framework for multivariate responses to model the joint distribution of two or more responses in the spirit of GAMLSS relying on a fully Bayesian approach.

% In this work, we develop an approach for fitting multivariate distributional regression models in the context of statistical boosting. 
An alternative approach to penalized regression and Bayesian approaches is statistical boosting, %which is a flexible alternative to classical estimation approaches 
which was originally developed in the field of machine learning and later extended to statistical modelling~\citep{friedman2000,friedman2001}. Its main features are the great flexibility regarding the effect types (e.g., spatial, smooth, or random effects) and the data-driven variable selection mechanism. The latter can be especially useful when the focus is on obtaining sparse models for a possibly high-dimensional covariate space~\citep{buhlmann2007}. %In particular, boosting for statistical modelling can deal with data situations in which the number of predictors exceeds the number of observations ($p > n$) and in which several classical approaches are no longer feasible~\citep{buhlmann2007}. 
The concept of boosting has already been extended to distributional regression leading to an algorithm that is able to estimate and select additive predictors for all distribution parameters in univariate GAMLSS~\citep{Mayr2012, Thomas2018}.

In this work, we adapt the boosting algorithm for multivariate responses by combining the properties of GAMLSS and the main features of statistical boosting. 
Due to the structure of the algorithm, our approach is able to simultaneously model all distribution parameters and to select possible predictor effects in multivariate distributional regression models: The new multivariate boosting approach allows to model not only the marginals but also the association between multiple outcomes through an additive predictor.

Motivated by three biomedical applications, we focus on modeling and investigating specific bivariate regression models with emphasis on common parametric distributions in biomedical research: the bivariate Bernoulli distribution for binary outcomes, the bivariate Poisson distribution for count data and the bivariate Gaussian distribution for continuous outcomes~\citep{MarshallOlkin1985,Kocherlakota1992,Kotz2000}.% and examine these in three different biomedical data situations.  % (\cite{MarshallOlkin1985} showed how other distributions can be naturally derived from the bivariate Bernoulli distribution.)

In the first biomedical application, the joint genetic predisposition for chronic ischemic heart disease and high cholesterol is analyzed based on a large cohort data from the UK Biobank~\citep{sudlow2015ukbiobank} via the bivariate Bernoulli distribution. The main interest is to study the dependence of these phenotypes on the genetic variants and to discover possible joint associations of the two outcome variables, which is not feasible via classical approaches modeling the phenotypes separately~\citep{MR2015}. In our case, we want to gain deeper insights into the relationship between the two phenotypes and the genetic variants affecting their association.

In the second application, we investigate effects for the demand on health care in Australia reported by \cite{Cameron1998} based on data from the Australian health survey. The two considered outcomes are the number of consultations with a doctor and the number of prescribed medications, whose association is modeled using the bivariate Poisson distribution for the covariates gender, age and annual income. The research question is based on a previous analysis by~\cite{KarlisNtzoufras2005}, however we illustrate that our approach offers higher flexibility. 

In the last epidemiological application, two indicators for undernutrition, namely for acute and chronic undernutrition, of children in Nigeria are jointly analyzed, which is motivated by a previous analysis by~\cite{KleKneKlaLan2015}. The two scores are modeled with a bivariate Gaussian distribution, in which besides the marginal expectations also the scale parameter and the correlation parameter depend on covariates. In addition to several covariates describing the life situation of the children, the mother and the household they are living, spatial effects based on regional information are incorporated. % It is of interest to investigate how the standard deviation, but also the correlation change depending on covariates.

% Whereby its from Interaction between acute and chronic undernutrition, detect predictor variables that inflcuended the association 

% https://citeseerx.ist.psu.edu/viewdoc/download?doi=10.1.1.857.307&rep=rep1&type=pdf

The structure of this article is as follows: Section~\ref{Method} starts with a brief introduction to multivariate distributional regression models. Then we investigate the different bivariate regression models and give an insight into statistical boosting with a description of the extended algorithm.    
In Section~\ref{Simulations} we illustrate different data settings using a simulation study while Section~\ref{Application} illustrates the application on biomedical research questions for the considered distributional regression models in Section~\ref{Method}.

\section{Boosting Multivariate Distributional Regression}\label{Method}

\subsection{The notion of multivariate distributional regression models}

In multivariate structured additive distributional regression~\citep{KleKneKlaLan2015} it is assumed that the conditional distribution $\dsP_{\mY|\mX=\xvec}$ of a $D$-dimensional vector of responses $\mY=(Y_1,\dots, Y_D)^\top$ given covariate information summarized in $\mX=\xvec$ has a $K$-parametric density $p(\yvec\mid\xvec)=p(\yvec\mid{\thetavec}(\xvec))$ with covariate dependent distribution parameters $\thetavec(\xvec)\equiv\thetavec=(\theta_1, \ldots, \theta_K)^\top$. 

Each distribution parameter $\theta_k$ is linked to a structured additive predictor $\eta_k$~\citep{FahKneLan2004} via bijective parameter-specific link functions $g_k$, such that $g_k(\theta_k)=\eta_k$ and $g_k^{-1}(\eta_k)=\theta_k$, $k=1,\ldots,K$. The inverse link functions $g_k^{-1}\equiv h_k$ are called response functions and ensure potential restrictions of the parameter space of $\theta_k$. The additive predictors $\eta_k$ depend on (possibly different) subsets of $\xvec$ and are of the form 
%\begin{equation*}
%    g_k(\theta_k) = \eta_k = \beta_{0k} + \sum\limits_{j = 1}^{p_k} f_{jk}(\xvec_{jk}),     \mbox{ for } k = 1,\dots, K,
%\end{equation*}
%where $\beta_{0k}$ are the intercepts and each $f_{jk}$, $j=1,\ldots,p_k$, represents the functional effect of covariate subset $\xvec_{jk}\subset\xvec$ on parameter $\theta_k$.
\begin{equation*}
    g_k(\theta_k) = \eta_k = \beta_{0k} + \sum\limits_{j = 1}^{p_k} f_{jk}(\xvec),     \mbox{ for } k = 1,\dots, K,
\end{equation*}
where $\beta_{0k}$ are the intercepts and each $f_{jk}$, $j=1,\ldots,p_k$, represents the functional effect of covariates $\xvec$.
The effects of the covariates can be specified in a very flexible manner and can correspond to linear, non-linear, random, interaction and further effects \citep[see e.g.,][]{FahKneLanMar2013,Woo2017}. Motivated by our applications in Section~\ref{Application}, in this work we focus on the following effect types:
\begin{enumerate}
    \item Linear effects are represented by $f_{jk}(\xvec) = \xvec_{jk}^T \betavec_{jk}$, where $\betavec_{jk}$ are the regression coefficients and $\xvec_{jk}$ is a covariate subset of $\xvec$ for parameter $\theta_k$ ($\xvec_{jk}$ can be chosen individually for each parameter $\theta_k$).
    \item Non-linear effects can be included using smooth functions $f_{jk}(\xvec)$. As basis functions we use B-Splines with second order difference penalties \citep{Eilers1996}.
    \item Spatial effects based on observations assigned to discrete regions are incorporated using Markov random fields for modeling neighborhood structures $f_{jk} (\xvec) = f_{jk}(s_i)$, where $s_i$ denotes the region $s_i$ observation $i$ is located in \citep{RueHel2005}.
\end{enumerate}

\subsection{Examples of relevant response distributions}\label{Distributions}

In the following, we describe three common bivariate parametric distributions for binary, count and continuous responses, representing the most common response types in biomedical research.  We will focus on the bivariate Bernoulli, the bivariate Poisson and the bivariate Gaussian distribution. While there are of course other multivariate distributions for discrete and continuous data \citep{Johnson1997,Kotz2000}, these three bivariate distributions are arguably most commonly used and are also relevant for our applications. 

\subsubsection{Bivariate Bernoulli distribution}

For analyzing potentially correlated binary variables $\mY  = (Y_1, Y_2)^T$, we consider the bivariate Bernoulli distribution with joint probability mass function
\begin{equation*}
    p(y_1,y_2) = p_{00}^{(1-y_1)(1-y_2)} p_{10}^{y_1(1-y_2)}
    p_{01}^{(1-y_1)y_2} p_{11}^{y_1y_2},~~ y_1,y_2\in\{0,1\},
\end{equation*}
where $p_{ij} = P(Y_1 = i, Y_2 = j)$, $i,j \in \{ 0,1 \}$ are the joint probabilities. Then, the contingency table with marginal probabilities $p_d = \mathrm{P}(Y_d = 1)$, $d = 1,2$ is given by:

\begin{table}[h!]
\centering
\begin{tabular}{l|lllr}
                       &   & \multicolumn{2}{c}{$Y_2$} &       \\ \midrule
                       &   & 0     & 1                &       \\ \midrule
 \multirow{2}{*}{$Y_1$} & 0 & $p_{00}$ & $p_{01}$        &    $1 - p_1$    \\
                       & 1 & $p_{10}$ & $p_{11}$            & $p_1$  \\  \midrule
                       &   &   $1 - p_2$      & $p_2$             &  1    
\end{tabular}
%\caption{Contingency table}\label{ConTab}
\end{table}

In a bivariate logistic regression model (logit model), the marginal probabilities $p_1 = \mathrm{P}(Y_1 = 1)$ and $p_2 = \mathrm{P}(Y_2 = 1)$, as well as the odds ratio $ \psi = \frac{p_{00} p_{11}}{p_{01}p_{10}}$ describing the association between the two binary outcomes, can be estimated considering several covariates \citep{mccullagh1989generalized,palmgren89}. If $Y_1$ and $Y_2$ are independent, then the odds ratio $\psi = 1$. The different additive predictors in the bivariate logit model are
\begin{align*}
    \logit(p_i) &=\eta_{p_i}, \mbox{ for } i = 1,2 \quad\mbox{ and }\quad\log(\psi)  = \eta_{\psi}.
\end{align*}
The joint probability $p_{11}$ can be determined from the marginal probabilities $p_1,p_2$ and the odds ratio $\psi$ via
\begin{equation*}
     p_{11} = \begin{cases}
     \frac{1}{2}(\psi - 1)^{-1}\lbrace a - \sqrt{ a^2 + b}\rbrace & \,,\, \psi \neq 1 \\
     p_1 p_ 2& \,,\, \psi = 1,
     \end{cases}
\end{equation*}
where $a = 1 + (p_1 + p_2)(\psi -1)$ and $b = -4 \psi (\psi-1) p_1 p_2$ \citep{Dale1986}. The joint probabilities $p_{10}, p_{01}$ and $p_{00}$ can be derived from $p_{11}$ and the marginal probabilities.

An alternative approach for modeling bivariate binary responses is the bivariate probit model. However, in this work we focus on the logit model for two reasons: First, one distribution parameter directly corresponds to the odds ratio, which is  easier to interpret and much more common in Biostatistics and biomedical research than the correlation of a latent bivariate response $\mY^\ast\sim N(\nullvec,\mSigma)$ for a probit model, where $Y_d=1\mbox{ if } Y_d^\ast>0 \mbox{ and } 0 \mbox{ otherwise, }d=1,2$ and $\mSigma$ a correlation matrix. Second, in a boosting and frequentist framework, the bivariate logit model is computationally favorable since it does not require the latent variables $\mY^\ast$.

\begin{comment}
\citep{AshfordSowden1970}
\begin{equation*}
    \my^{*} = \mathbf{\eta}_\mu + \epsilon, \quad \epsilon \sim N(\mathbf{0}, \mathbf{\Sigma}),
\end{equation*}
where $\mathbf{y^*} = (y^*_1, y_2^*)^T$ is an unobserved latent variable, which follows a multivariate Gaussian distribution with $\mathbf{\mu} = (\mu_1,\dots, \mu_D)^T = (\mathbb{E}(y^*_1), \dots, \mathbb{E}(y^*_D))^T$ and covariance $\mathbf{\Sigma}$. The covariance matrix has values of 1 on the diagonal and on the off-diagonal $\text{corr}(d_1,d_2)$ with $d_1,d_2 = 1,\dots,D$.
Furthermore, the latent variable is positive if $y=1, y^*>0$.
\end{comment}
%\begin{align*}
%     p_{11} &= \begin{cases}
%     \frac{1}{2}(\psi - 1)^{-1}\lbrace a - \sqrt{ a^2 + b}\rbrace & \psi \neq 1 \\
%     p_1 p_ 2& \,\psi = 1
%     \end{cases} \\
%     p_{01} &= p_2 - p_{11}\\
%     p_{10} &= p_1 - P_{11}\\
%     p_{00} &= 1- p_{11} - p_{01} - p_{10} = 1 + p_{11} - p_1 -p_2
%\end{align*}
%with $a = 1 + (p_1 + p_2)(\psi -1)$ and $b = -4 \psi (\psi-1) p_1 p-2$

\subsubsection{Bivariate Poisson distribution}\label{BP}

% \textbf{is there a bivariate negative binomial distribution?}\\
% https://www.cambridge.org/core/journals/annals-of-actuarial-science/article/application-of-bivariate-negative-binomial-regression-model-in-analysing-insurance-count-data/C8A1A76A40C8D1EA2BA9CADD13F3D55B 
% Has different parameter specification

An important bivariate model for analyzing bivariate count data can be constructed from combining three random variables. If $Z_k, k = 1, 2, 3$ follow independent Poisson distributions with parameters $\lambda_k > 0$, then the two random variables $Y_1 = Z_1 + Z_3$ and $Y_2 = Z_2 + Z_3$ follow a bivariate Poisson distribution with joint probability function given by
\begin{equation*}
    p(y_1, y_2) = \exp{(-(\lambda_1 + \lambda_2 + \lambda_3))} \frac{\lambda_1^{y_1}}{y_1 !} \frac{\lambda_2^{y_2}}{y_2 !} \sum\limits_{k = 0}^{\min{(y_1,y_2)}} \binom{y_1}{k} \binom{y_2}{k} k! \left(\frac{\lambda_3}{\lambda_1 \lambda_2}\right)^k, ~~ y_1,y_2\in\mathbb{N}_0.
\end{equation*} 
The marginals also follow Poisson distributions with expectations $\mathbb{E}(Y_1) = \lambda_1 + \lambda_3$ and $\mathbb{E}(Y_2) = \lambda_2 + \lambda_3$. The parameter $\lambda_3$ controls the dependency between $Y_1$ and $Y_2$ and  corresponds to the covariance $\mathrm{Cov}(Y_1,Y_2) = \lambda_3$. If the variables $Y_1$ and $Y_2$ are independent, then $\lambda_3 = 0$ and the bivariate Poisson distribution reduces to the product of two independent Poisson distributions. For further details on the bivariate Poisson distribution, see \cite{Kocherlakota1992} and \cite{Johnson1997}.

In a bivariate Poisson  model, each distribution parameter $\lambda_k,\, k=1,2,3$ can be modeled in terms of several explanatory variables via
\begin{align*}
 \log(\lambda_k) = \eta_{\lambda_k},k=1,2,3,
\end{align*}
where $\eta_k$ is the corresponding  predictor for $\lambda_k$. 
% By modeling the covariance~$\lambda_3$, one can obtain detailed insights into the influence of different covariates on the joint distribution~\citep{KarlisNtzoufras2005}. %The marginal mean for $Y_1$ can be expressed, for example, by $\mathbb{E}(Y_{1}) = \exp(\xvec_{1}^{T}\beta_1) + \exp(\xvec_{3}^{T}\beta_3)$ and for $Y_2$ accordingly with $\mathbb{E}(Y_2) = \exp(\xvec_{2}^{T}\beta_2) + \exp(\xvec_{3}^{T}\beta_3)$, with covariate subset $\xvec_{k} = (x_{1k},\dots, x_{p_k k})^T$, $k = 1,2,3$ for the distribution parameters \citep{KarlisNtzoufras2005}.

A drawback of this definition of the bivariate Poisson distribution is its property of modeling only data with positive correlations. An alternative was developed in \cite{Lakshminarayana1999} by defining the bivariate Poisson distribution as the product of Poisson marginals with a multiplicative factor. This definition also allows for negative correlations, but results in more difficult interpretations. A further alternative allowing for overdispersion in the marginal distributions is the bivariate negative binomial distribution~\citep{Kocherlakota1992,Ma2020}. We refrain from describing this distribution in more detail given our application in Section~\ref{HealthCare}, where previous works have considered the bivariate Poisson distribution to be a reasonable modeling choice~\citep{KarlisNtzoufras2005}. In general, however, our approach of course would be also feasible for alternative parameterizations or alternative distributions.

% Deeper insight and short discussion: \cite{Famoye2010}

%  Chib and Winkelmann (2001), van Ophem 1999, Berkhout and Plug 2004 - Allows negative correlatino but much more complicated and require special efforts for parameter estimation
% http://www2.stat-athens.aueb.gr/~karlis/multivariate%20Poisson%20models.pdf

%(More flexible from by Lakshminarayana: positive, negative correlation
%bivariate distribution whose marginals are Poisson developed as a product of Poisson marginals with a multiplicative factor. But in the case of a boundary value of the dependence parameter (i.e. $\lambda_3$ assuming a very small negative value), the model might fail to adequately fit the data and may provide a false positive scenario.(A new bivariate Poisson distribution via conditional specification: properties and applications - Ghosh et al. 2020))

%same covariance for all pairs - which is unrealistic, comments and alternative version for covariance: \url{https://link.springer.com/content/pdf/10.1007/s11222-005-4069-4.pdf}, in general limited: complexity of calculating the probability distribution function}).\\

\subsubsection{Bivariate Gaussian distribution}\label{BivNormal}
The bivariate Gaussian distribution is one of the most commonly known distributions for considering two continuous responses. In this case, the random vector is written by $\mY \sim N(\muvec,\mSigma)$, where the density of $\mY= (Y_1, Y_2)^T$ is given by
\begin{equation*}
  f(y_1,y_2) = \frac{1}{2\pi\sqrt{\mathrm{det}( \mSigma)}}\exp\left(-\frac{1}{2} (\yvec-\muvec)^{T}\mSigma^{-1}(\yvec-\muvec)\right),~~ y_1,y_2\in\mathbb{R},
\end{equation*}
and $\muvec = (\mu_1, \mu_2)^T$ and $\mSigma = \Cov(Y_1,Y_2)$ are its mean vector and covariance matrix, respectively. The latter is defined by
\begin{equation*}
    \mSigma = \begin{pmatrix} \sigma_1^1 & \rho \sigma_1 \sigma_2 \\  \rho \sigma_1 \sigma_2  & \sigma_2^2\end{pmatrix}
\end{equation*}
with marginal variances $\sigma^2_1 = \Var(Y_1)$ and $\sigma^2_2 = \Var(Y_2)$ and correlation parameter $\rho = \Cor(Y_1,Y_2)$. 
All parameters of the bivariate Gaussian distribution can be again modeled depending on covariates with parameter specific link-functions:
\begin{align*}
 \mu_1 = \eta_{\mu_1}, \quad  \mu_2 = \eta_{\mu_2}, \quad
 \log(\sigma_1) = \eta_{\sigma_1}, \quad \log(\sigma_2) = \eta_{\sigma_2} \quad \mbox{and} \quad \rho/\sqrt{(1-\rho^2)} = \eta_{\rho}.
\end{align*}
For further practical and theoretical details of the bivariate Gaussian distribution, we refer to \cite{Kotz2000}.

When the marginal distributions exhibit heavy tails, the bivariate $t$-distribution is an attractive alternative to the bivariate normal distribution. Motivated by our application on childhood undernutrition in Section~\ref{Undernutrition}, where the normality assumptions for the considered scores is reasonable~\citep[see e.g][]{KleCarKneLanWag2021}, we omit details on the bivariate $t$-distribution in the context of structured additive distributional regression and refer to \citet{KleKneKlaLan2015} and references therein for further modeling details.

\subsection{Estimation via model-based boosting}

Boosting originally arose from the field of supervised machine learning~\citep{Freund1990} but gained increasing popularity in statistics after the concept was adapted to fit statistical regression models~\citep{friedman2000,friedman2001}. Boosting algorithms are a flexible alternative to classical estimation approaches and have several practical advantages, such as the applicability to high-dimensional data problems and data-driven variable selection~\citep{buhlmann2007,Li2022,Wu2019}. 
In the context of regression, there exist different types of boosting algorithms~\citep{Tutz2006,buhlmann2007}. Here, we will focus on a component-wise gradient boosting algorithm with regression-type base-learners, which we refer to as \textit{statistical boosting}~\citep{Mayr2014,Mayr2014extending}. 

This statistical boosting approach is based on minimizing a pre-specified loss function, which represents the regression problem and typically corresponds to the negative log-likelihood $l$ of the response distribution. 
In every iteration of the boosting algorithm, so-called base-learners are separately fitted to the negative gradient of the loss function, and the best-performing one is updated to the current estimate. A base-learner in our context is a regression function, and usually corresponds to one specific covariate effect in the additive predictor (e.g., a linear model as base-learner leads to a linear effect). An overview of possible base-learners can be found in~\cite{Hofner2014}.
% Finally, the best performing base-learner is updated in every boosting step. 

\begin{figure}[t]\centering
    \includegraphics[width=\textwidth]{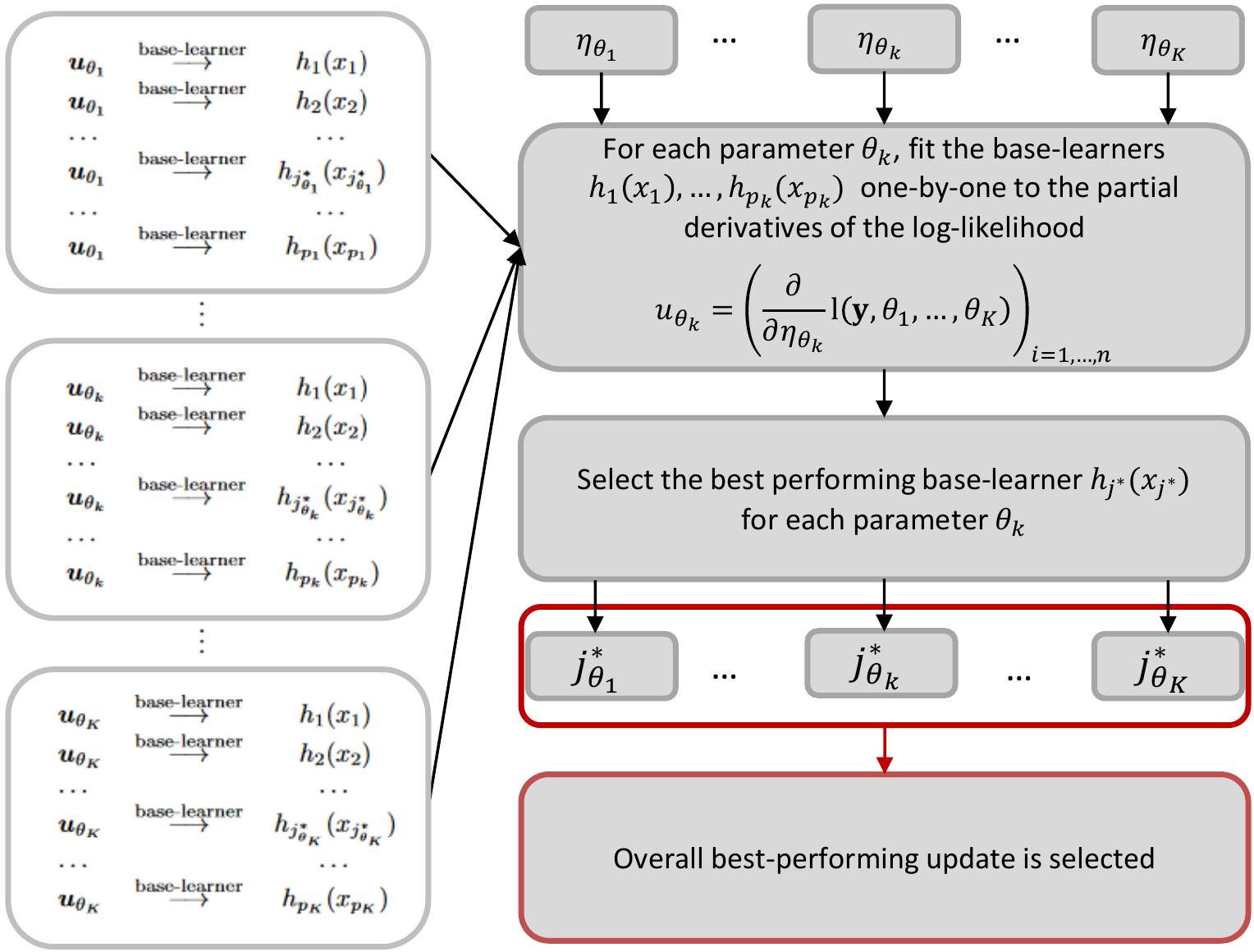}
    \caption{Graphical representation of boosting multivariate structured additive distributional regression (displaying one boosting iteration).}\label{FlowChart}
\end{figure}

For fitting multivariate distributional regression models, we extend the statistical boosting algorithm for generalized additive models for location, scale and shape~\citep{Mayr2012} to multivariate distributions.
A schematic overview of the selection of base-learners in one iteration of  the boosting algorithm for multivariate responses can be found in Figure~\ref{FlowChart}.

First, for each additive predictor $\eta_{k}$, $k=1,\dots,K$, a set of base-learners $h_{1}(x_{1}), \dots, h_{p_k}(x_{p_k})$ has to be specified in advance. Then, the partial derivative $u=\partial l/\partial\theta_k$  of the negative log-likelihood function $l$ with respect to the different distribution parameters $\theta_k$ is calculated and each base-learner is fitted separately to the gradient of the corresponding parameter $k$. For each parameter, the best performing base-learner  $j_{k}^*$ is determined.
After these best-fitting base-learners are selected for each dimension $k$, only the overall best update (with the highest loss reduction) of all distribution parameters is finally added to the corresponding additive predictor, with the estimated effect multiplied by a small fixed step-length, e.g., $\nu = 0.1$. 
That means, in every iteration the best-fitting base-learner is determined for each distribution parameter and then compared across the different dimensions. This refers to a so-called non-cyclic version of boosting for distributional regression, leading to a single update of only one distribution parameter in each iteration~\citep{Thomas2018}.%(Scalar-Optimization: Partitioning of base-learner between different parameters is done automatically while fitting the model. Do not need to optimize the stopping iteration for each distribution parameter.)

The main tuning parameter of the algorithm is the number of boosting iterations, which is typically chosen by cross-validation or resampling techniques. As the algorithm is usually stopped before convergence (\textit{early stopping}), the optimization of the stopping iteration leads to the prevention of overfitting and encourages the sparsity of the resulting model by data-driven variable selection~\citep{MayrHofnerSchmid2012}. In particular, those variables, whose corresponding base-learners have never been selected in the update process, are effectively excluded from the final model. The variable selection is simultaneously based on all additive predictors of the corresponding multivariate distribution. The algorithm does not impose any hierarchy between distribution parameters, but only judges the potential predictor variables based on their performance in increasing the joint likelihood. In addition, early stopping typically leads to an improvement in the prediction accuracy and shrinkage of the effect estimates. We provide an implementation of statistical boosting for multivariate distributional regression,  which is integrated in the \textsc{R} package \textbf{gamboostLSS} \citep{Hofner2016}. 
% We provide an implementation of statistical boosting for multivariate distributional regression, in an add-on package gamboostLSS.
%The corresponding \textsc{R} code to reproduce the results is available on \textcolor{red}{GitHub https://github.com/AnnikaStr/DistRegBoost} 

\section{Simulations}\label{Simulations}
To evaluate the performance of the proposed statistical boosting approach, we conducted a detailed simulation study for the three response distributions presented in Section~\ref{Distributions}. 
For each distribution, the particular settings are guided by the different applications in Section~\ref{Application}. With our simulations, we aim to answer the following questions:
\begin{itemize}
    \item Does the boosting approach yield accurate estimates for the corresponding distribution parameters of the bivariate distributions?
    \item Can the boosting approach identify the truly informative variables and their effects?
    \item How do the bivariate models perform compared to univariate models that assume independence between the two response components?
\end{itemize}
In particular, we evaluate the estimation, variable selection and predictive performance. Note that for each considered simulation setting, different variables are informative for the distribution parameters and some of them partially overlap. Therefore, we refer to informative and non-informative variables and do not mention all of them individually for the different settings. 

For all simulations, the step-length (learning rate) of the boosting algorithm is set to a fixed value of $\nu = 0.1$ for each parameter of the bivariate models, as well as for the univariate boosted models. This is currently common practice in statistical boosting~\citep{SCHMID2008298,Mayr2012,Hofner2014}. The stopping iteration $m_{\rm{stop}}$ is optimized by minimizing the empirical risk on an additional validation data set with $n_{\text{val}} = 1500$ observations, following the same distribution as the training data. 
In addition, test data with 1000 observations were generated for the evaluation of the predictive performance (from the same distribution as the training data). A total of 100 simulation runs were performed for each simulation setting. The corresponding \textsc{R} code to reproduce the results is available on GitHub \url{https://github.com/AnnikaStr/DistRegBoost}.

%All simulations were conducted in the statistical computing software \textsc{R}~\citep{R}. 

% Further simulation results can be found in Appendix
% the same conditions applied to the univariate boosted models

\subsection{Bivariate Bernoulli distribution}
\subsubsection{Simulation design}
For the simulation of the bivariate logit model, we considered a high-dimensional setting with $n = 1000$ observations and $p= 1000$ covariates for each of the three parameters. For data generation, the \textsc{R} package \textbf{VGAM}~\citep{VGAM} was used, whereby the parameters $p_1, p_2$ and $\psi$ were simulated with the following linear predictors
\begin{align*}
    \logit(p_1)  = \eta_{\mu_1} = X_1 + 1.5 X_2 - X_3 + 1.5 X_4,\qquad 
    & \logit(p_2)  = \eta_{\mu_2} = 2X_1 - X_2 + 1.5 X_3, \\
    \log(\psi)   = \eta_{\psi} = - 1.5 +& 1 X_5 + 1.5 X_6.
\end{align*}
Overall, only the first six covariates out of the $p = 1000$ had a relevant effect on any of the distribution parameters (four for $p_1$, three for $p_2$ and two for $\psi$). The covariates were simulated from a multivariate normal distribution $N(\nullvec,\mSigma)$ with a Toeplitz covariance structure $\Sigma_{ij} = \rho^{|i-j|}$ for $1\leq i,j\leq p$, where $\rho =0.5$ is the correlation between consecutive variables $X_j$ and $X_{j+1}$.  
The covariates were incorporated in the boosting approach by using simple linear models as base-learners.
As measures for the predictive performance, the area under the curve (AUC), the Brier score, the negative log-likelihood and energy score were considered. Note that the AUC and Brier score do not account for the dependence between the two outcomes and are calculated separately for both outcomes.

\begin{figure}[t]\centering
    \includegraphics[width=\textwidth]{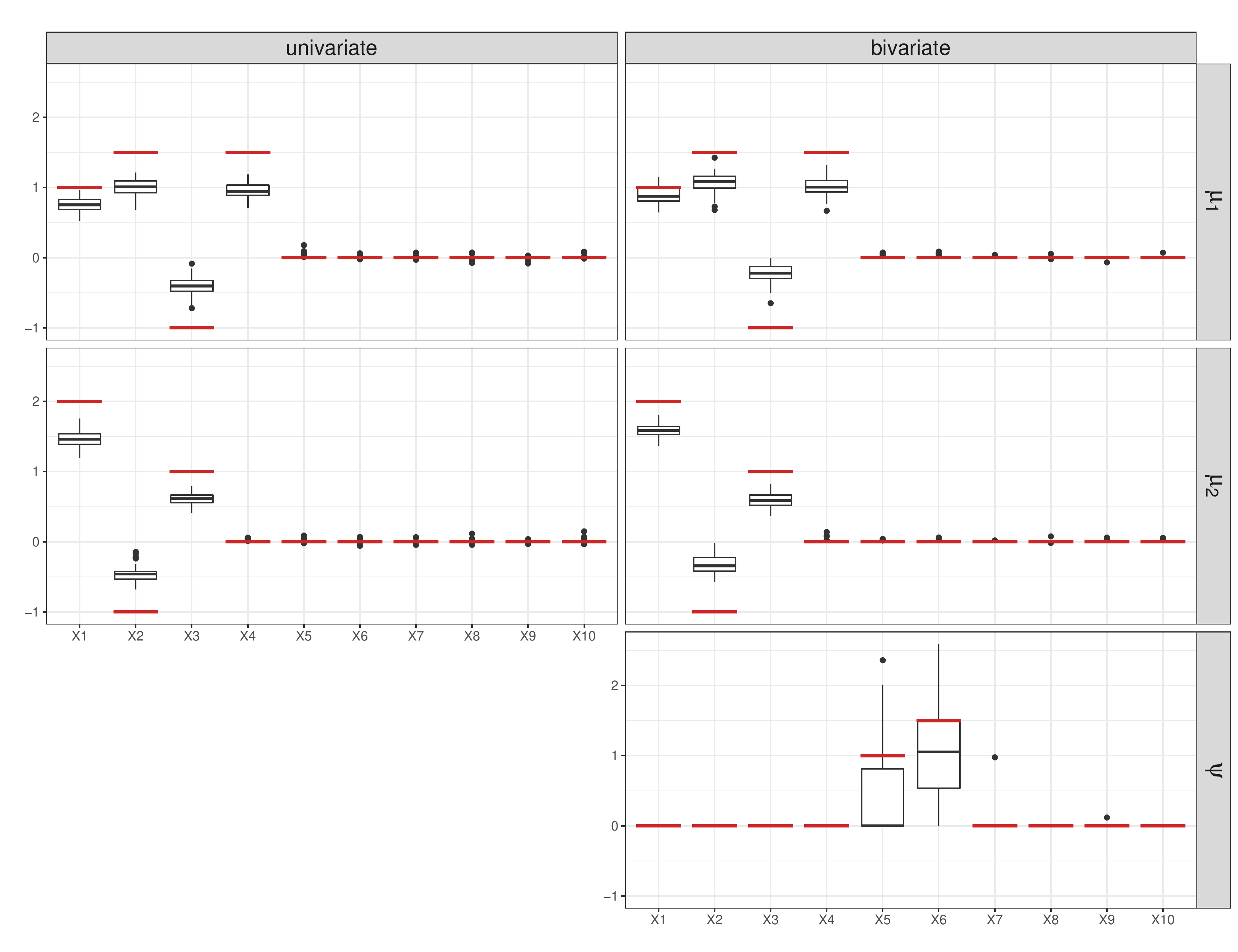}
    \caption{Results for the estimated linear effects of the univariate (left) and bivariate Bernoulli (right) model of the first ten covariates $X_1, \dots, X_{10}$ from 100 simulation runs. The red horizontal lines correspond to the true values.}\label{Figure:Bern1}
\end{figure}

\subsubsection{Results}
Figure~\ref{Figure:Bern1} presents the coefficient estimates of the first ten covariates $X_1, \dots, X_{10}$ in form of boxplots resulting from 100 simulation runs for the univariate (left) and bivariate (right) model with the red horizontal lines corresponding to the true values. 
The univariate and bivariate models reflect the true structure for $\eta_{\mu_1}$ and $\eta_{\mu_2}$, as well as $\eta_{\psi}$ for the bivariate model, with both models leading to very similar results. 
The informative variables for $\mu_1$ and $\mu_2$ were almost selected in each simulation run, leading to an overall selection rate (average value over the informative variables) of 100\% for the univariate models for both parameters and 100\% for $\mu_1$ and 97.75\% for $\mu_2$ in the bivariate model. The selection rate for $\psi$ is a bit lower than for the other parameters with a selection rate of 59\% (see Appendix~Table~A3). The non-informative variables were selected very rarely overall, resulting in sparse models. %that highlight the variable selection properties of gamboostLSS.
A comparison of the predictive performance is provided in Table~\ref{Table:Logit}, showing that the univariate and bivariate models were very similar in terms of AUC, Brier score, and energy score, with the bivariate model having slightly better negative log-likelihood. In addition, the energy score for the univariate models showed a larger standard deviation. 
Further simulation results of this linear setting for a low-dimensional data situation ($p=10$ and $n=1000$) can be found in Appendix~A.1.

\begin{table}[t]
    \centering
    \begin{tabular}{lcc}
    \toprule
         & Univariate & Bivariate \\ \midrule
        AUC ($Y_1$) & 0.88 (0.01) & 0.88 (0.01)  \\ 
        AUC ($Y_2$) & 0.85 (0.01) & 0.84 (0.01)  \\ 
        Brier score ($Y_1$) & 0.14 (0.01) & 0.14 (0.01) \\ 
        Brier score ($Y_2$) & 0.16 (0.01)  & 0.16 (0.01) \\ 
        Energy score & 0.25 (0.19) & 0.27 (0.01) \\ 
        Negative log-likelihood & 929.85 (22.46) & 906.68 (27.32) \\ \bottomrule
    \end{tabular}
    \caption{Resulting predictive performance on independent test data for the linear setting of the bivariate Bernoulli distribution; mean (sd) values from 100 simulation runs are reported for the univariate and bivariate models.}\label{Table:Logit}
\end{table} % Engergy Score: results are more robust and considers the tails of the distribution.

\subsection{Bivariate Poisson distribution}
\subsubsection{Simulation design}
For the bivariate Poisson regression model, we investigated both linear and non-linear settings with $p=10$ covariates and $n= 1000$ observations for each distribution parameter.
For the linear setting, the  underlying true predictors were specified as
\begin{equation}
\begin{aligned}\label{PoisLin}
    \log(\lambda_1)  = \eta_{\lambda_1} =  -X_1 + 0.5X_2 + 1.5 X_3,  \qquad
    & \log(\lambda_2)  = \eta_{\lambda_2} = 2X_1 - X_3 + 1.5 X_4 + X_5,  \\
    \log(\lambda_3)  = \eta_{\lambda_3} = 0.5 X_5 &+ X_6 - 0.5 X_7, 
\end{aligned}
\end{equation}
where the covariates followed a multivariate normal distribution $N(\nullvec,\mSigma)$ with Toeplitz covariance structure and correlation coefficient $\rho =0.5$. Thus, the first seven covariates were informative for any of the distribution parameter (three for $\lambda_1$ and $\lambda_3$, four for $\lambda_2$). For this setting, simple linear models were incorporated as base-learners. 
For the non-linear setting, the true additive predictors were given by
\begin{equation}
\begin{aligned}\label{PoisNonlin}
    \log(\lambda_1)  = \eta_{\lambda_1} = \sqrt{X_1}X_1,
   \qquad &\log(\lambda_2)  = \eta_{\lambda_2} = \cos(2X_2), \\
    \log(\lambda_3)  = \eta_{\lambda_3} &= \sin{X_3}, 
\end{aligned}
\end{equation}
where the covariates were independently simulated from the uniform distribution~$U(0,1)$ and only one covariate was informative for each of the distribution parameters. As base-learners, we chose P-splines (20 equidistant knots with a second-order difference penalty and four degrees of freedom). 
The \textsc{R} \textbf{extraDistr} package from \cite{Wolodzko2020} was used to simulate data from the bivariate Poisson regression model.

\begin{figure}[t]\centering
    \includegraphics[width=\textwidth]{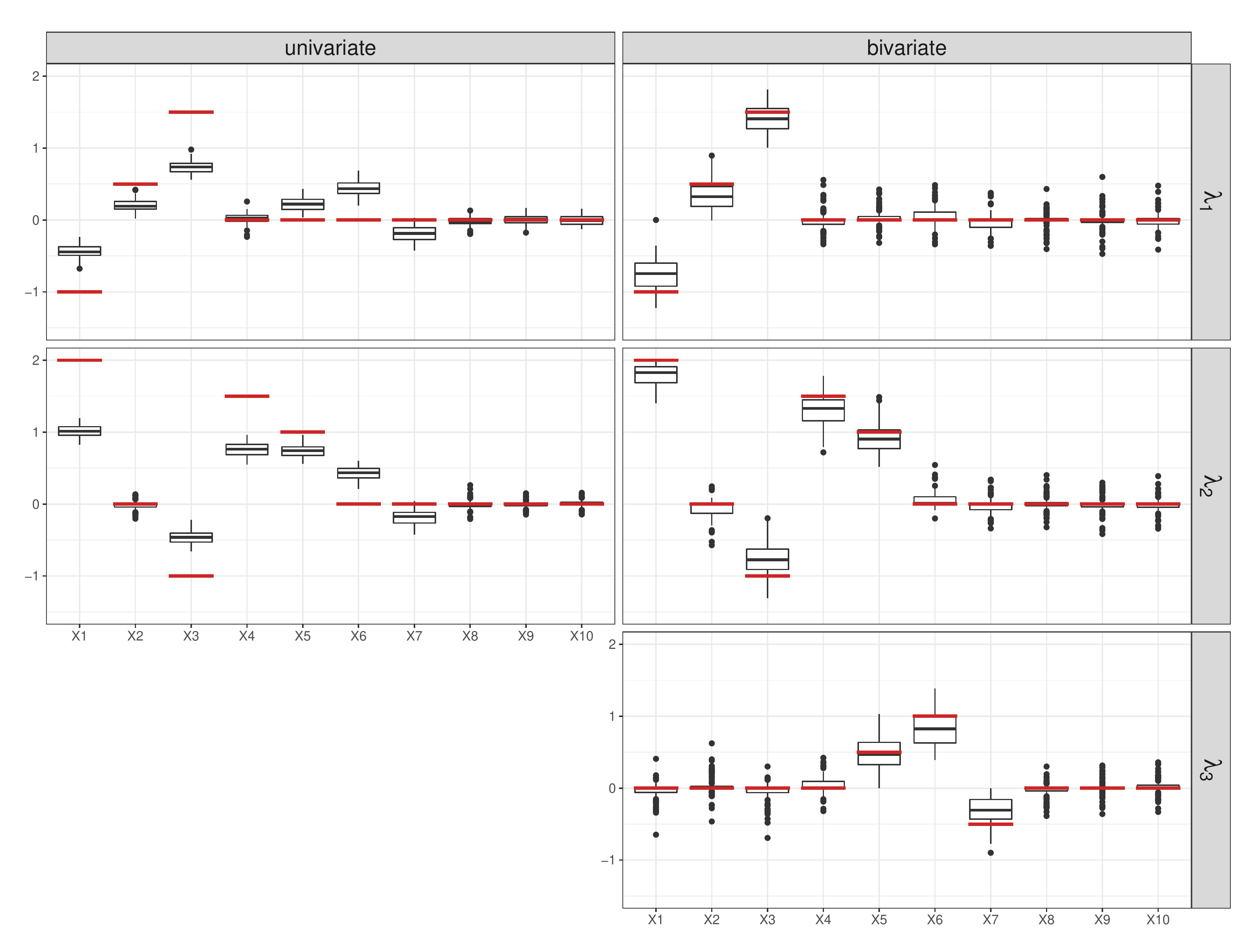}
    \caption{Results for the estimated linear effects of the univariate (left) and bivariate Poisson model (right) from 100 simulation runs. The horizontal lines correspond to the true values.}\label{Figure:Pois1}
\end{figure}

\subsubsection{Results}
Figure~\ref{Figure:Pois1} displays the coefficient estimates for the linear Poisson regression models~\eqref{PoisLin}. The boxplots present the estimated coefficients of the 100 simulation runs for the univariate (left) and bivariate models (right). % the empirical distribution   
Overall, boosting the bivariate regression model was able to identify the informative variables and to accurately estimate the true effects represented by the red horizontal lines. In comparison, the univariate models for $\lambda_1$ and $\lambda_2$ resulted in much smaller estimated coefficients.
For both models, the informative variables were selected in almost every simulation run: considering $\lambda_1$ and $\lambda_2$, the univariate models and the bivariate model had a selection rate of almost 100\%
for the informative variables, whereby also for $\lambda_3$ a high selection rate of 95.67\% for the informative variables was achieved.  
On the other hand, the univariate models as well as the bivariate model selected also several non-informative variables with a small coefficient size. A more detailed overview on the selection rates for the specific parameters can be found in Table~A4 of the Appendix.

Furthermore, we considered the MSEP, the negative log-likelihood, and the energy score for the evaluation of the predictive performance on test data (see Table~\ref{Table:Pois}). The MSEP only accounts for the marginal distributions and displays here a slightly better performance for the univariate models. The negative log-likelihood and the energy score, which also take the association into account, showed a better performance for the bivariate model. %and had a higher standard deviation of the energy score for the univariate models. 
\begin{figure}[t]\centering
    \includegraphics[width=\textwidth]{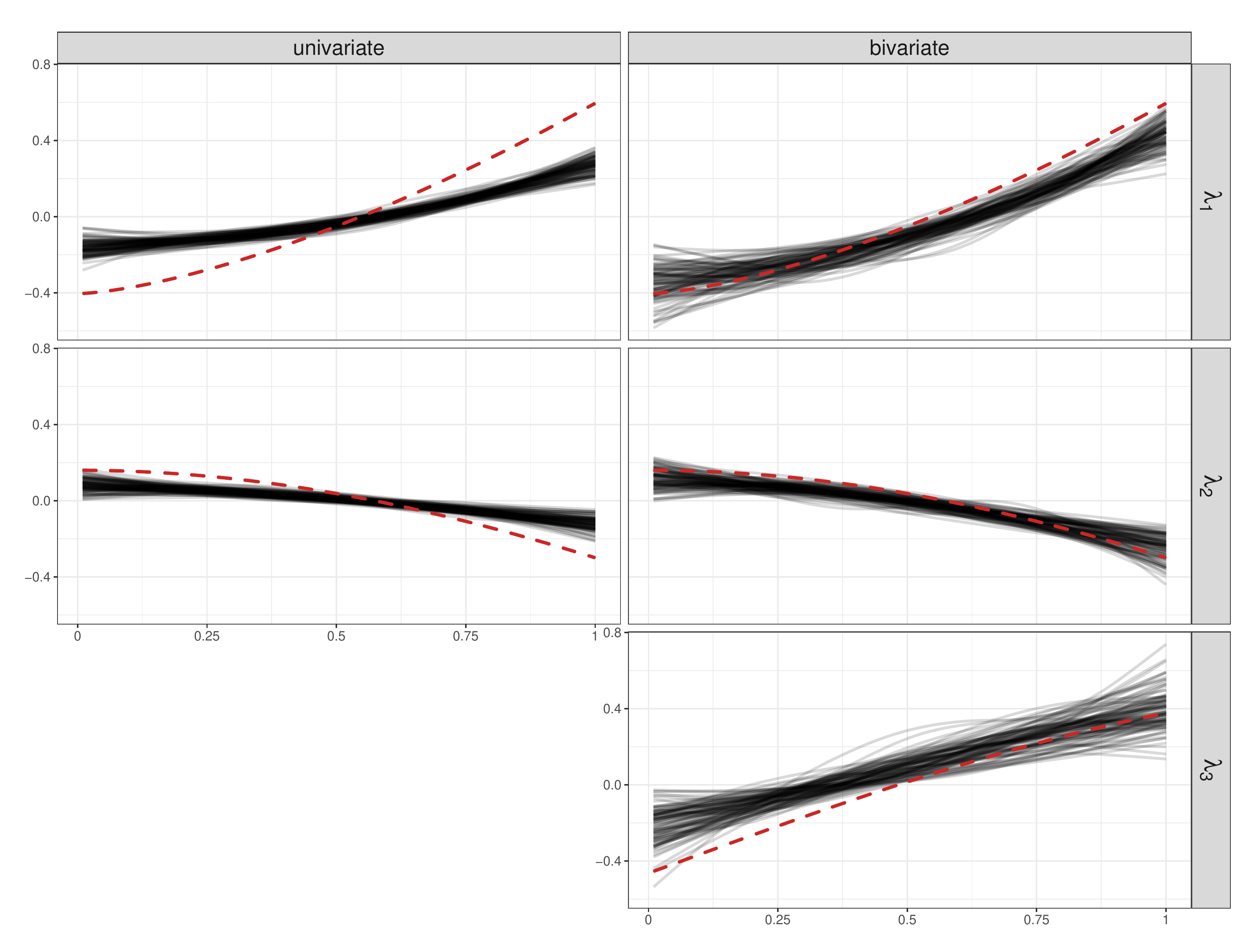}
    \caption{Results for the estimated non-linear effects for the univariate (left) and the bivariate Poisson model (right) from 100 simulation runs. The red dotted lines correspond to the true effects.}\label{Figure:Pois2}
\end{figure}

Figure~\ref{Figure:Pois2} displays the effect estimates for the non-linear setting~\eqref{PoisNonlin}. %Plotted are the estimated functions from the 100 simulation runs for the informative variables (black) along with the true underlying effects (red dotted lines). 
Overall, the estimated splines approximate the true effects well for each parameter of the bivariate model and clearly outperform the univariate models for $\lambda_1$ and $\lambda_2$.
%For the univariate model, identifying the true functions was more problematic, and the splines shrank more toward zero. 
The informative variables were selected in each simulation run. However, as in the linear model, we observed also high selection rates for the non-informative variables in both models (see Appendix~A.2). % for more details).  

In terms of predictive performance, similar to the linear setting, the MSEP indicated a better performance of the univariate models, while the bivariate models outperformed the univariate models in terms of the negative log-likelihood as expected. The energy score is very similar for both models but slightly better for the bivariate model overall.
Further simulation results for these settings in case of high-dimensional data with $p=1000$ covariates and $n = 1000$ observations can be found in Appendix~A.2.
% showing that selection rates lower, 

%The informative variables $X_1, X_2$ and $X_3$ had a selection rate of 100\% for the respective parameter (selected in each simulation run). 
% Overall the the splines approximated the true function well (how good could the algorithm detect the true function, effect estimates are shrunk towards zero )
% Selection rates ... 

\begin{table}[t]
    \centering
    \resizebox{\textwidth}{!}{%
    \begin{tabular}{l|cc|ccc}
    \toprule
        &\multicolumn{2}{c|}{Linear model} &  \multicolumn{2}{c}{Non-linear model}\\
        &  Univariate & Bivariate & Univariate & Bivariate \\ \midrule
        MSEP ($Y_1$) & 2.66 (0.18) & 3.96 (0.29) & 4.64 (0.25) & 8.18 (0.65) \\
        MSEP ($Y_2$) & 2.86 (0.23) & 4.11 (0.34) & 5.49 (0.29) & 9.06 (0.68) \\ 
        Energy score & 1.48 (1.11) & 1.36 (0.03) & 1.95 (0.04) & 1.95 (0.04) \\ 
        Negative log-likelihood  & 3598.42 (54.31) & 3413.68 (40.91) & 4433.06 (52.06) & 4246.96 (42.58) \\\bottomrule
    \end{tabular}}
    \caption{Resulting predictive performance on independent test data for the linear and non-linear settings of the bivariate Poisson regression; mean (sd) values from 100 simulation runs are reported for the univariate and bivariate models.\label{Table:Pois}}
\end{table}

\subsection{Bivariate Gaussian distribution}
\subsubsection{Simulation design}
For the simulation of a bivariate Gaussian distributed outcome, we considered a setting with linear, non-linear and spatial effects with $p=10$ covariates and $n=1000$ observations with the following true predictors
\begin{small}
\begin{align*}
    \mu_1  = \eta_{\mu_1} = \sin(2X_1)/0.5 + X_6 + 0.5 X_7 + f_{\text{spat}} \qquad &
    \mu_2  = \eta_{\mu_2} = 2 + 3\cos(2X_2) + 0.5X_7+X_8 + f_{\text{spat}}\\
    \log(\sigma_1) = \eta_{\sigma_1} = \sqrt{X_3}X_3 -0.5X_8 + f_{\text{spat}} \qquad &
    \log(\sigma_2) =\eta_{\sigma_2} = \cos(X_4)X_4 + 0.25X_9 + f_{\text{spat}}\\
    \rho/\sqrt{1-\rho^2} = \eta_{\rho} = \log(X_5^2) &+ X_{10}+ f_{\text{spat}},
\end{align*}
\end{small}
where the covariates were independently simulated from the uniform distribution~$U(0,1)$. Each included covariate was informative for one of the distribution parameters; more precisely, for each parameter three covariates, one linear and one non-linear, and additionally the spatial effect. For linear effects we used simple linear models as base-learners and P-splines for the non-linear effects. 
The spatial effects were simulated with $f_{\text{spat}}(s) = \sin(x^{c}_s)\cos(0.5y^{c}_s), s \in {1,\dots,S}$, based on the centroids of the standardized coordinates of the discrete regions in Western Germany with overall $S=327$ regions. % (available \url{https://www.uni-goettingen.de/de/550530.html}). 
The neighborhood structure was modeled by the spatial base-learner using a Markov random field \citep[based on the \texttt{R} package \textbf{BayesX} by][]{Umlauf2019}. 
%Here, the mean squared error of prediction (MSEP), the negative log-likelihood, and the energy score were considered as measures of the predictive performance.

%As with the bivariate Bernoulli and Poisson distribution, the number of boosting iterations is validated by an additional data set with $n_{\text{val}} = 1500$ from the same distribution as the training data set and the step-size was chosen with 0.1. For the evaluation of the predictive performance, a test data set with 1000 observation was generated under the same condition as the training data. 

\begin{table}[t]
    \centering
    \begin{tabular}{lcc}
    \toprule
         & Univariate & Bivariate \\ \midrule
        MSEP ($Y_1$) & 1.59 (0.11) & 1.59 (0.11) \\ 
        MSEP ($Y_2$) & 1.38 (0.07) & 1.38 (0.07) \\ 
        Energy score & 1.03 (0.02) & 1.01 (0.02) \\ 
        Negative log-likelihood & 3370.41 (89.59) & 3098.11 (109.97) \\ \bottomrule
    \end{tabular}
    \caption{Resulting predictive performance on independent test data of the bivariate Gaussian regression; mean (sd) values from 100 simulation runs are reported for the univariate and bivariate models.\label{Table:Norm}}
\end{table}

\subsubsection{Results}
\begin{figure}[h]\centering
    \includegraphics[width=\textwidth]{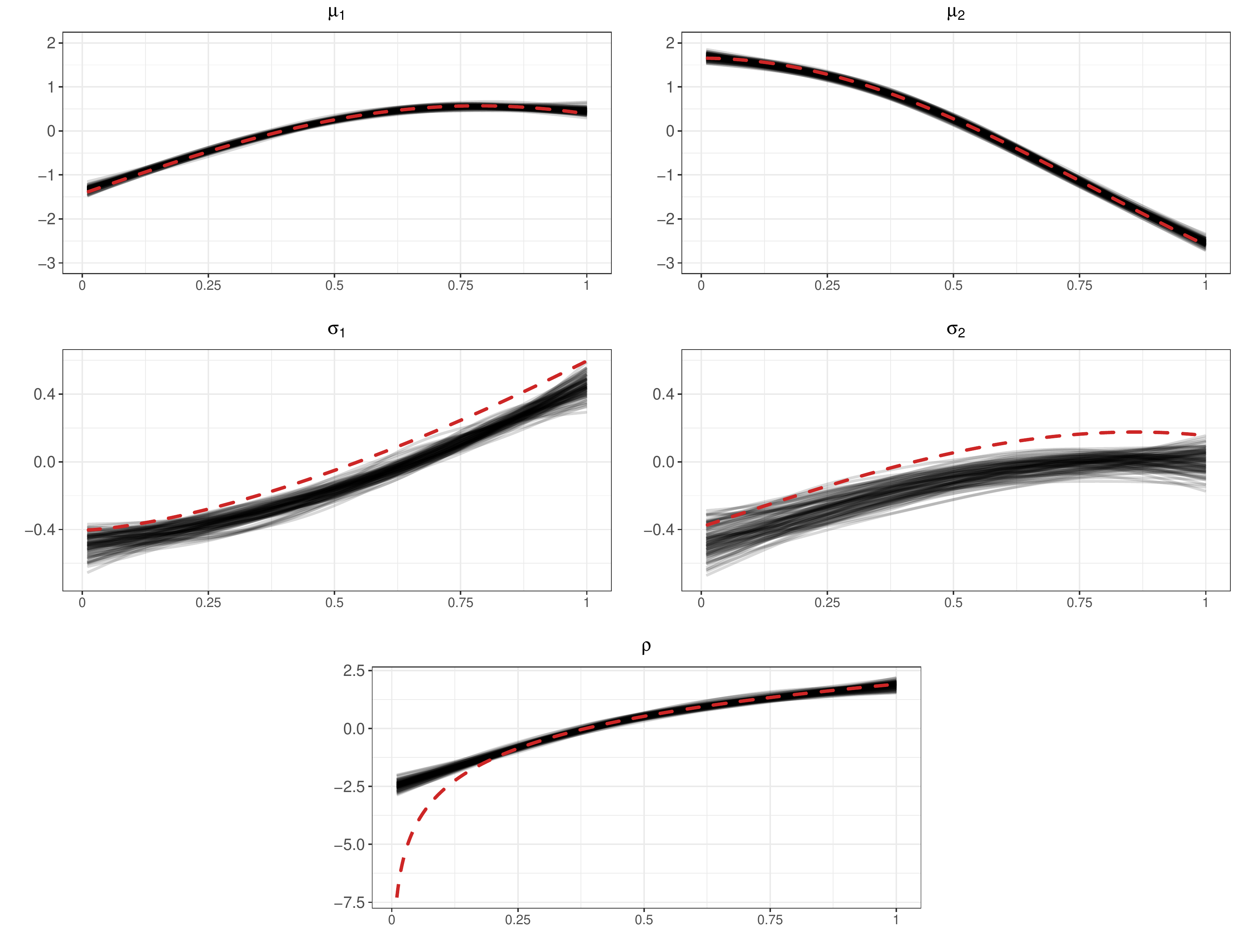} 
    \caption{Results for the estimated non-linear effects of the bivariate Gaussian regression model from 100 simulation runs. The red dotted lines correspond to the true effects.} \label{Fig:Normal}
\end{figure}

Considering the linear effects (see Figure~A4 in Appendix~A.3), the effect estimates from the 100 simulation runs for both models reflect the true structures of the linear part of the predictors, whereby the bivariate model better approximates the true values. The bivariate model was also able to capture the true non-linear functions well (%red dotted lines in 
Figure~\ref{Fig:Normal}); only small deviations are observed for the variance and for the correlation~$\rho$ at the left border. The results for the univariate models appear to be very similar regarding the univariate effects and can be found in the Appendix~(Figure~A5). 
For the spatial effects, the true structure for the regions in West Germany was identified by each distribution parameter (a graphical representation of the true structure and the estimated spatial effects are in Appendix~A.3).

The informative variables for the univariate and bivariate models were selected in nearly all 100 simulation runs, where the bivariate model also correctly selected the informative variables for the correlation between the outcomes. Whereas, we can not examine the correlation with the univariate models. The selection rates for the non-informative variables were slightly higher for the bivariate model (see Appendix~Table~A7). %A detailed overview can be found in Table~A6 of the Appendix.

Regarding predictive performance, the MSEP, the energy score, and the negative log-likelihood were considered. For the MSEP and the energy score, similar results were observed for the univariate and the bivariate models. The negative log-likelihood on the test set showed an improvement in predictive performance considering the bivariate model.
Further simulation results for this setting in case of high-dimensional data with $p=1000$ covariates and $n = 1000$ observations can be found in Appendix~A.3.

\subsection{Summary} % Summary
Overall, we obtained promising results for all three considered distributional regression families (logistic, Poisson and Gaussian regression), highlighting that the boosting algorithm yields appropriate estimates for the different parameters and is capable of identifying the most informative variables from a potentially much larger set of candidate variables.  
The comparison with the univariate models showed that the estimated effects for the bivariate model was able to provide better approximations to the true structure of the predictors than the univariate models (particularly for the Poisson and Gaussian regression models). 
We noticed that for the logistic regression model, the selection rates for the association parameter, the odds ratio, tended to be lower than for the association parameter of the bivariate Poisson and Gaussian distribution. 
Furthermore, the number of selected non-informative variables was higher for the univariate as well as the bivariate models for the Poisson distribution. However, the linear low-dimensional setting for the bivariate logistic regression model (see Appendix~A.1) and the low-dimensional Gaussian regression model showed higher selection rates in this situation as well (Appendix~A.3). Conclusively, this highlights a tendency of the algorithm to select more non-informative variables in low-dimensional settings. % Same for the low-dimensional linear Bernoulli setting in Appendix, and for the Normal distribution the selection rates are not as high as for the Poisson, but also higher

Regarding prediction accuracy, as expected, the univariate and bivariate models performed similarly for evaluation criteria that consider only the marginals (AUC, brier score and MSEP). Only for the Poisson distribution, the univariate model performed slightly better regarding the MSEP. This can be explained by the particular design of this bivariate distribution, i.e.\ the summation of the means for both outcomes ($\mathbb{E}(Y_1) = \lambda_1 + \lambda_3$ and $\mathbb{E}(Y_2) = \lambda_2 + \lambda_3$). In Figure~\ref{Figure:Pois1}, for example, we observe that the informative variables $X_5, X_6$ and $X_7$ for parameter $\lambda_3$ were selected quite frequently with a higher estimated coefficient in the univariate models. These wrongly selected variables for the marginals resulted in an improvement of the MSEP. In the bivariate model, we account for the association between $Y_1$ and $Y_2$ by modeling the dependency in terms of the covariates. The MSEP does not account for the association and the variables describing dependency are not reflected in the marginals as in the univariate models. %Therefore, the MSEP for the bivariate model is slightly higher. 
 
%  univariate model selects the informative variables of $\lambda_3$ the MSEP can be better
Regarding the predictive scores which account for associations between the outcomes, the energy score tended to be very similar for the univariate and bivariate models, while the negative log-likelihood was consistently better for the bivariate models. %Both the energy score and the negative log-likelihood account for the association between the outcomes.

\section{Biomedical applications}\label{Application}

In this section we consider three diverse biomedical data sets to illustrate the applicability of our extended boosting approach for multivariate distributional regression models based on binary, count and continuous outcomes presented in Section~\ref{Distributions}.

\subsection{Genetic predisposition for chronic ischemic heart disease and high cholesterol}
% https://www.dge.de/wissenschaft/weitere-publikationen/fachinformationen/niedriges-ldl-und-hohes-hdl-cholesterol-senken-das-risiko-fuer-kardiovaskulaere-ereignisse/
% > 6.19: High

For analyzing the association between high cholesterol and chronic ischemic heart disease in dependency of different genetic variants, we used cohort data from the UK Biobank. The UK Biobank is a large biomedical cohort study containing genetic and health information from over half a million British participants~\citep{sudlow2015ukbiobank}.

In classical approaches for analyzing a potential genetic liability to a specific phenotype such as high cholesterol or chronic heart disease, each considered genetic variant is fitted individually to the phenotype using a simple linear model~\citep{MR2015}. 
In this context, previous works including genome-wide association studies~\citep[e.g.,][]{LinselNitschke2008,Richardson2020} have investigated to find genetic variants associated with high cholesterol and heart disease.
Using our boosting algorithm for multivariate distributional regression, the main interest here is to investigate the association between chronic ischemic heart disease and high cholesterol, both considered as binary phenotypes (high cholesterol $>$ 6.16 mmol/l). In particular, we aim to identify genetic variants affecting their association by estimating the two phenotypes jointly in a bivariate logistic  model. That means we do not only want to model the individual distributions of the two phenotypes, but also estimate the dependency between these phenotypes as a function of genetic variants, which is not possible with conventional approaches. 

The considered data set consists of 20,000 randomly sampled observations of individuals with white British ancestry, with  additional 10,000 observations used to validate the optimal stopping iteration. 
For each phenotype, 1000 variants were selected in a pre-screening step based on the largest marginal associations between the variants and the phenotype, which were computed with the PLINK2 function \texttt{-variant-score}~(\citealt{Plink2, ChangPlink}). After pre-screening, the data set contains a total of 1865 variants (with 135 variants selected for both phenotypes). Variants with minor allele frequency not less than 1\% were randomly sampled with the\texttt{-thin-count} function. Missing genotypes were imputed by the reference allele using the \texttt{R} package \textbf{bigsnpr}~\citep{Priv2018}.

\begin{figure}[h]\centering
    \includegraphics[width=\textwidth]{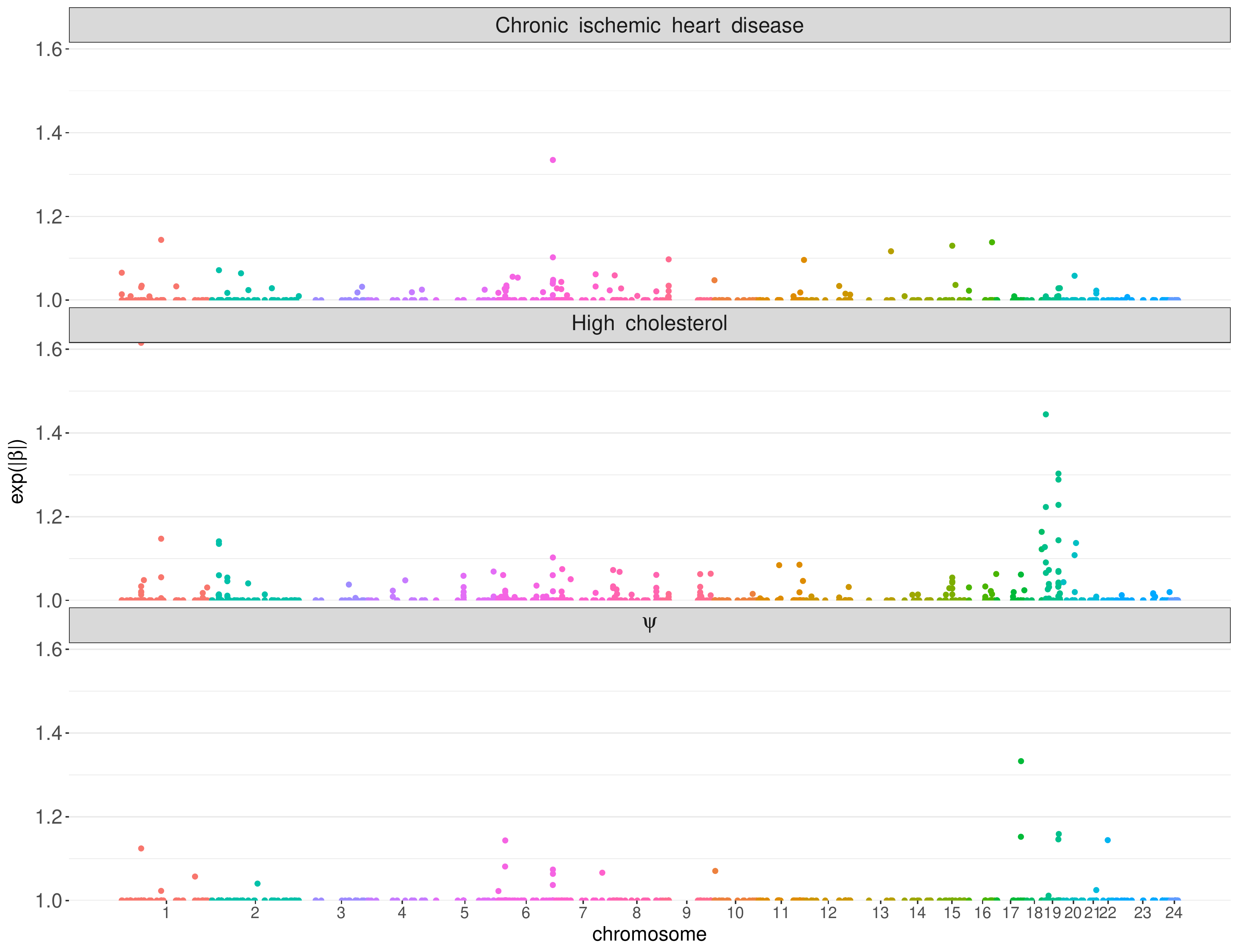} 
    \caption{Manhattan plots for the coefficient estimates (expressed in exponential absolute values of the estimated coefficients) of the boosted bivariate logistic regression model of the joint analysis of high cholesterol and ischemic heart disease from the UK Biobank data. The $x$-axis represents the genomic location of the variants.} \label{ManhattanPlot}
\end{figure} % Lücken sind dadurch entstanden, das entweder dort nicht gemessen wurde oder diese Variante nicht in meinem subset war.

Figure~\ref{ManhattanPlot} shows the resulting estimated coefficients  (expressed in exponential absolute values of the estimated coefficients) for the three distribution parameters. 
When comparing these Manhattan plots with the classical univariate ones (based on the marginal association evaluated on the $-\log10(P)$ scale) for high cholesterol and chronic ischemic heart disease, we find that the bivariate boosting model tended to identify variants with a higher coefficient value (stronger effect) from similar or the same genomic locations, where the univariate models also showed large univariate associations (see Appendix~B.1).

For high cholesterol, e.g., the variants with the smallest univariate p-values are located on chromosomes 18 and 19; on these chromosomes there were also the variants that had the highest estimated coefficients in the bivariate boosting model. These findings are consistent with the location of known cholesterol-associated genes~\citep{Richardson2020}. Variants from these chromosomes were also selected with our approach for the odds ratio. Our model selected several variants for chronic ischemic heart disease that are in line with the findings of the meta-analysis of genome-wide association studies examining DNA sequence variants associated with ischemic heart disease of \citet{ELOSUA2017754} (e.g., the variants rs11206510, rs2891168 and rs4420638).

Overall, several variants were selected by the boosting approach for each distributional regression parameter, i.e., 75 for $\mu_1$, 154 for $\mu_2$, and 19 variants for $\psi$. For each parameter, mainly those variants were selected that were primarily filtered due to the specific phenotype (the 1000 most highly associated from the univariate screening). 
In particular, for $\mu_1$, 63 of the 75 selected variants had been primarily chosen for ischemic heart disease, so that 12 of the 75 selected variants for $\mu_1$ also had a large univariate association with high cholesterol. Regarding high cholesterol, 110 variants were selected from the ones that had been pre-selected, while 44 of the selected variants for high cholesterol had been originally pre-selected for ischemic heart disease.
The two marginal means $\mu_1$ and $\mu_2$ had six selected variants in common, and both had one variant that was also selected for the odds ratio $\psi$ (namely for $\mu_1$: rs10455872 and for $\mu_2$: rs77542162).
The odds ratio included two variants that were among the 1000 most highly correlated pre-selected variants for both phenotypes, namely rs505151 and rs2229094. The other 17 variants selected for the odds ratio were divided as follows: 10 from $\mu_1$ (ischemic heart disease) and 7 from $\mu_2$ (high cholesterol).
This means the algorithm identified several variants that affect the dependency between the two phenotypes. The odds ratio is the most common measure for examining the dependency between two binary outcomes in biomedical research and the interpretation in our context is very similar. Thus, the selected variants for the association parameter have an effect on both outcomes, with a positive effect increasing the association between heart disease and high cholesterol and conversely.
%For a detailed insight into the interpretation of the odds ratio we refer to \citet[Chapter 6.6.2][]{mccullagh1989generalized}. 

In summary, our algorithm provides the ability to study the joint genetic predisposition for chronic ischemic heart disease and high cholesterol. With our approach we can also model the dependence of the association between these two phenotypes on genetic variants, which is not possible with classical approaches. In addition, in line with the literature on cardiovascular genetics, our model selected several variants in genomic regions which had been previously identified to be relevant for the considered phenotypes. 

% Literature paper association of cholesterol and heart disease - Richardson2020
% https://www.ncbi.nlm.nih.gov/pmc/articles/PMC7089422/
% Analyzing lipoproteine with CHD (coronary heart disease) -> LDL cholesterol, triglycerides and apolipoprotein B had effect estimates consistent with a higher risk of CHD (individually MR)
% 

% Gibt auch Paper die das anders machen
% https://journals.plos.org/plosone/article?id=10.1371/journal.pone.0002986#amendment-0

% Paper: Meta-Analyse: Übersicht mit Genome-wide Association Studies Examining DNA Sequence Variants Associated With Ischemic Heart Disease
% https://www.revespcardiol.org/en-pdf-S188558571730289X
% Haben welche gemeinsam

%  PRS paper - variants for LDL-cholesterin
% one variant in common, chromosome 19; rs72658867, coefficient: 1.47

\subsection{Demand for health care in Australia}\label{HealthCare}
% features: compare to original paper and see which covariates effect which marginal and the dependence.
% data: from the bivpois on github (Package is not in CRAN anymore)
The first analysis on the demand for health care in Australia, based on the Australian health survey from 1977 to 1978, was reported by \cite{Cameron1998}. The considered data set consists of $n =5,190$ observations (which is only a subset of the overall collected survey). The bivariate count variables of interest are the number of consultations with a doctor (in the past 2 weeks) and the number of prescribed medications (used in the last two days), which we model using bivariate Poisson regression. The explanatory variables are \textit{gender} (female coded as 1, male as 0), \textit{age} (in years divided by 100) and annual \textit{income} (in Australian dollars; AUD;  divided by 1000, measured as midpoints of coded ranges). More details on the survey and its original analysis can be found in~\cite{Cameron1998}.
The data are provided in the \textsc{R} package \textbf{bivpois}~\citep{KarlisNtzoufras2005}, which is available on GitHub (\url{https://github.com/cran/bivpois}).

In the following, we use the same representation of the bivariate Poisson distribution as introduced in Section~\ref{BP}. Each distribution parameter $\lambda_k, k=1,2,3$ is modelled based on explanatory variables. We consider the two following models: %, where age and the income were included with non-linear effect.

\begin{enumerate}[leftmargin=*, label=Mod.~\Alph*,ref={Mod.~\Alph*}]
    \item\label{ModelA} \textit{Gender}, \textit{age} and \textit{income} are included as covariates for $\lambda_{\text{consulations}}$ (number of doctor consultations) and $\lambda_{\text{medications}}$ (number of medications prescribed), but only \textit{gender} is considered as a covariate for the covariance parameter $\lambda_3$ (corresponding to Mod.~(b) in \cite{KarlisNtzoufras2005}). 
    \item\label{ModelB} For each model parameter, P-splines are used as base-learners for the continuous variables \textit{age} and \textit{income}, while for \textit{gender} linear effects are used.
\end{enumerate}

\begin{table}[t]
\centering
\begin{tabular}{rlccrrr}
  \toprule
  & Covariate &  \multicolumn{1}{c}{$\lambda_{\text{consulations}}$} & \multicolumn{1}{c}{$\lambda_{\text{medications}}$} & \multicolumn{1}{c}{$\lambda_{3}$}  \\ 
  \midrule
 \ref{ModelA} &  Intercept & -2.10 & -2.20 & -2.62\\
            &  Gender (female) & 0.05 & 0.59 &  0.61 \\
            &  Age & 1.40 & 3.29 & -\\
            &  Income & -0.31 & -0.10 & - \\
  \midrule\ref{ModelB}&  Intercept & -2.29 & -2.22 & -0.35\\
            &  Gender (female) & 0.13 & 0.60 &  0.19\\
  %           &  Age & $f(\cdot)$ & f(\cdot) & f(\cdot)\\
  %          &  Income & f(\cdot) & f(\cdot) & f(\cdot) \\
 \bottomrule
\end{tabular}
\caption{Results of the bivariate Poisson model for the demand of Health Care for~\ref{ModelA} and~\ref{ModelB} (see Figure~\ref{Fig:HealthCare} for the non-linear effect estimates for age and income in \ref{ModelB}).\label{Tab:HealthCare}}
\end{table}

Considering the results of~\ref{ModelA} presented in the upper part of Table~\ref{Tab:HealthCare}, we observe that with increasing \textit{age}, both the numbers of doctor consultations and prescribed medications are estimated to increase. \textit{Income} has negative marginal effects on both responses, which means that higher \textit{income} is associated with fewer prescribed medications and fewer doctor appointments. For the covariance parameter $\lambda_3$, only \textit{gender} was included as an explanatory variable in~\ref{ModelA}. The joint effect of \textit{gender} on the number of doctor consultations and prescribed medications indicates that males and females have different covariance terms. The estimated effect of 0.61 for \textit{gender} suggests that the association between numbers of consultations and medications is higher for women than for men. % Results are can be compared with \cite{KarlisNtzoufras2005}.  
%   Higher association in females compared to males

\begin{figure}[t]\centering
    \includegraphics[width =\textwidth]{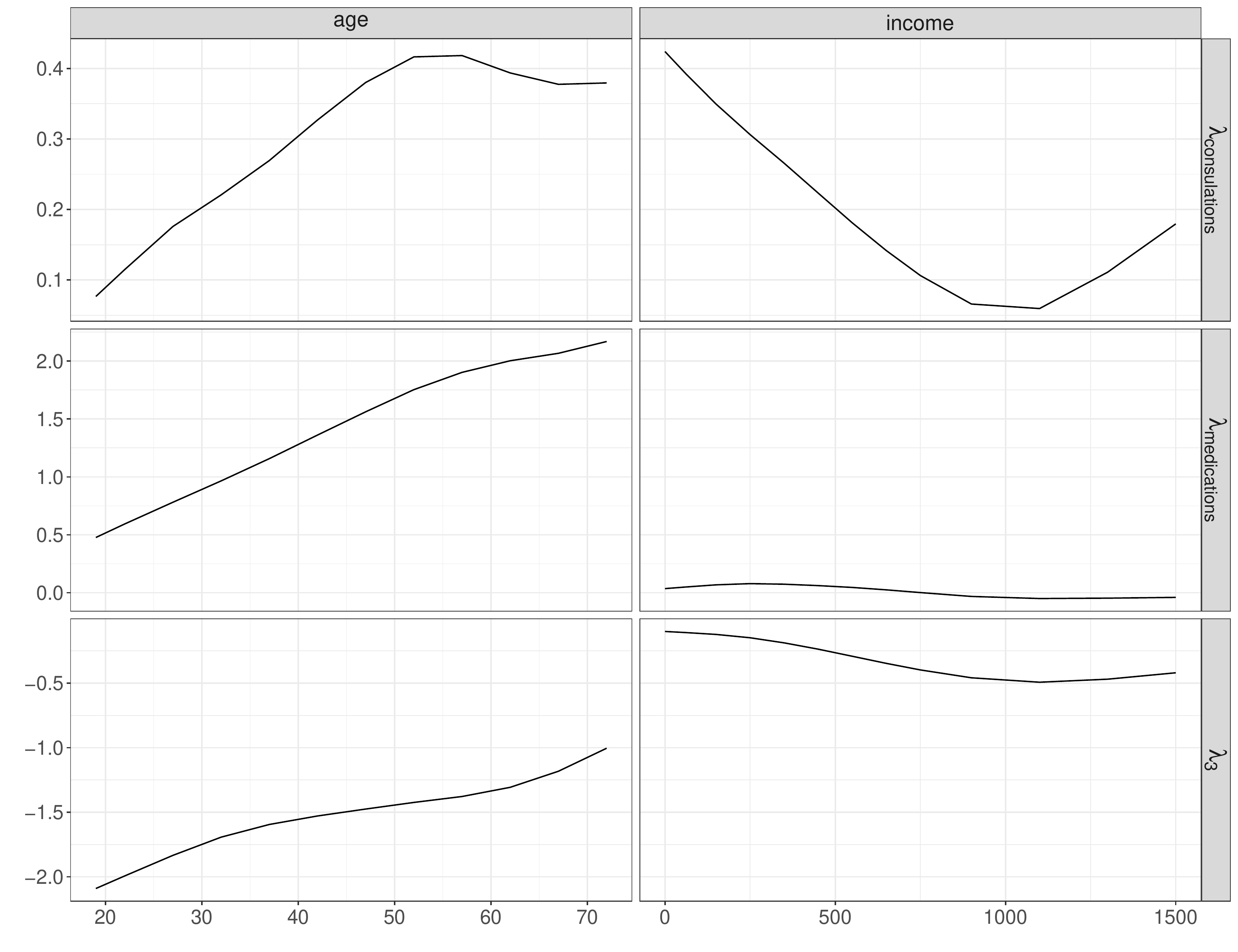}
    \caption{Partial effects of age and income on the demand for health care in Australia for~\ref{ModelB}.\label{Fig:HealthCare}}
\end{figure}

The lower part of Table~\ref{Tab:HealthCare} and Figure~\ref{Fig:HealthCare} present the results for~\ref{ModelB}. With an increasing \textit{age} up to 50 years, the number of doctor consultations is estimated to increase linearly, with a slight decrease starting around the age of 57 years. The estimated effect of \textit{age} on the number of prescribed medications and the covariance parameter is linear and is negative throughout the covariance. 
The \textit{income} is estimated to have a U-shape effect for the medical consultation, with a minimum between 800 AUD and 1,150 AUD. 
The estimated effect of \textit{gender} for \ref{ModelA} is slightly larger than for \ref{ModelB}.

Overall, the estimated effects of \ref{ModelA} are consistent with the results of \cite{KarlisNtzoufras2005}. In addition, we also considered a non-linear model. Both the linear and non-linear models indicated that the expected numbers of doctor consultations and prescribed medications increase with \textit{age}. For \textit{income}, the expected numbers of doctor visits and prescribed medications decreases with increasing \textit{income} for \ref{ModelA}. The expected number of doctor visits also decreases in \ref{ModelB} as \textit{income} increased, whereby a U-shaped effect for \textit{income} can be observed. % due to the non-linear estimation. 

Furthermore, because of the bivariate modeling, we also obtain information about the relationship between the outcomes. 
Here, both models showed a higher association between the number of doctor consultations and prescribed medications for women. Furthermore, \ref{ModelB} also included \textit{age} and \textit{income} as covariates for the covariance parameter and the model suggested that the association becomes greater with increasing \textit{age}.

% https://www.google.com/url?sa=t&rct=j&q=&esrc=s&source=web&cd=&ved=2ahUKEwjPicmMkavyAhVvgP0HHZy2DtMQFnoECAMQAQ&url=https%3A%2F%2Fwww.jstatsoft.org%2Farticle%2Fview%2Fv014i10%2Fv14i10.pdf&usg=AOvVaw0x_A2N8pGLbbkw8JbE9N0m
% http://sabre.lancs.ac.uk/exercise_c6.html

\subsection{Risk factors for undernutrition in Nigeria}\label{Undernutrition}   % https://arxiv.org/pdf/1902.10446.pdf

To analyze childhood undernutrition, a large database is available from the Demographic and Health Survey (DHS, \url{https://dhsprogram.com/}), containing nationally representative information about the population's health and nutrition status in numerous developing and transition countries. Here, we consider a data set used in \citet{KleCarKneLanWag2021} which contains data from Nigeria collected in 2013 with overall $23,042$ observations (after exclusion of outliers and inconsistent observations).
The bivariate responses are \textit{stunting}, which is defined as stunted growth measured as the insufficient height of the child concerning its age (chronic undernutrition), and \textit{wasting}, which refers to insufficient weight for height (acute undernutrition).
We analyze the joint distribution of these two responses using the bivariate Gaussian distribution with covariate-dependent marginal means and standard deviations as well as a covariate-dependent correlation parameter.

% \begin{equation*}
%  \binom{\text{stunting}}{\text{wasting}} \sim  N \left ( \binom{\mu_1}{\mu_2}, \begin{pmatrix} \sigma_1^2 & \rho \sigma_1 \sigma_2 \\  \rho \sigma_1 \sigma_2  & \sigma_2^2\end{pmatrix} \right)
% \end{equation*}

For continuous variables, P-splines were applied as base-learners, namely for \textit{cage} (age of the child in months), \textit{edupartner} (years of partner's education), \textit{mage} (age of the mother in years) as well as \textit{mbmi} (body mass index of the mother). Several other categorical covariates (12 covariates in total, e.g., \textit{bicycle}, \textit{car}, \textit{cbirthorder}) were included using simple linear models as base-learners. Furthermore, the neighborhood structure of the districts in Nigeria was incorporated and modeled by the spatial base-learner using a Markov random field. For a full description of the explanatory variables, see Appendix~B.2.
% Stunted growth is defined as a reduced growth rate compared to a standard population. Negative values of the score indicate that the child's growth is below the expected growth of a child with normal nutrion. 

\begin{figure}[t]
    \centering
    \includegraphics[width=\textwidth]{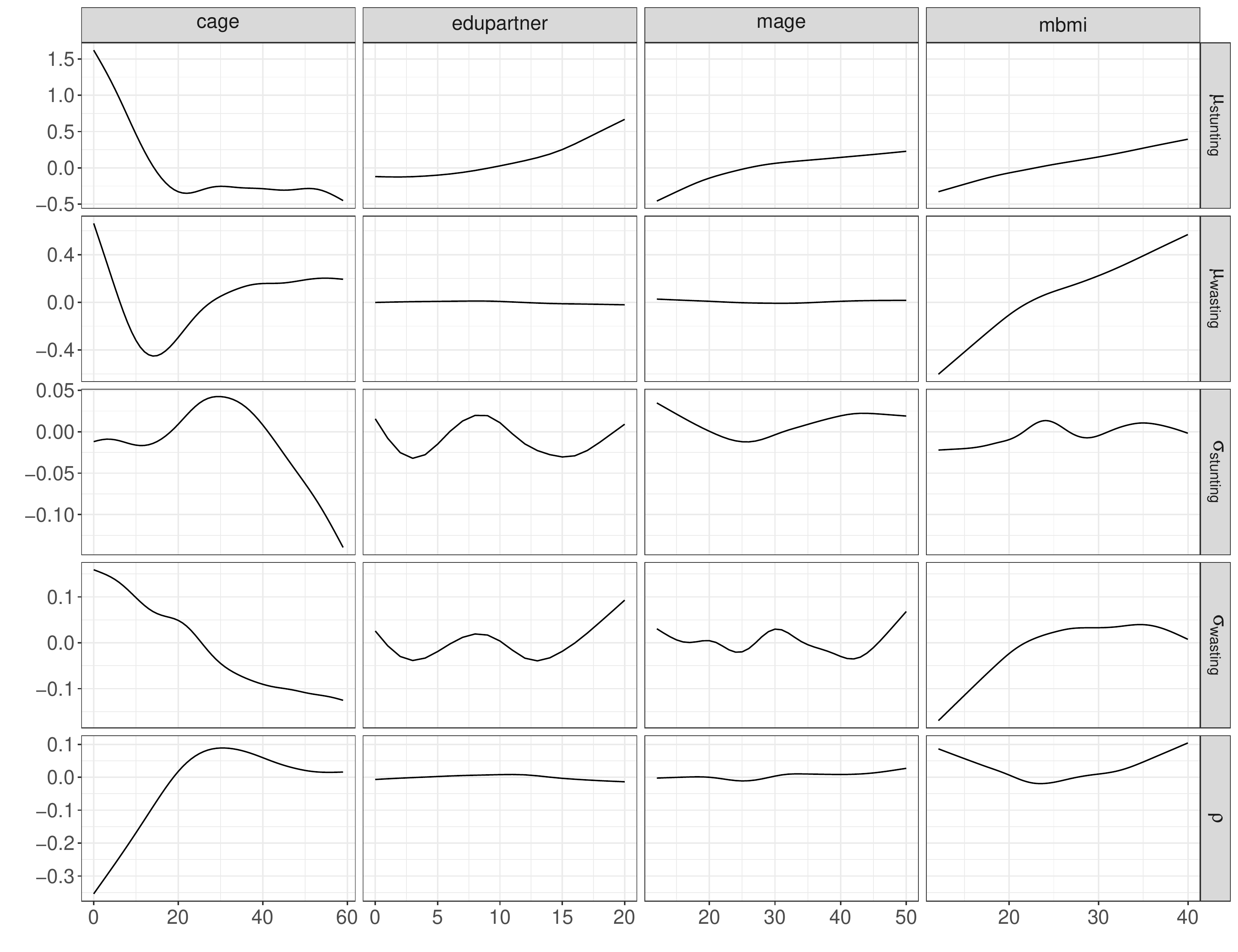}
    \caption{Non-linear effects of \textit{cage}, \textit{edupartner}, \textit{mage} and \textit{mbmi} for \textit{stunting} and \textit{wasting} of the bivariate Gaussian regression model for the Nigeria data.}\label{Fig:Nigeria1}
\end{figure}

\begin{figure}[t]\centering
    \includegraphics[width=\textwidth]{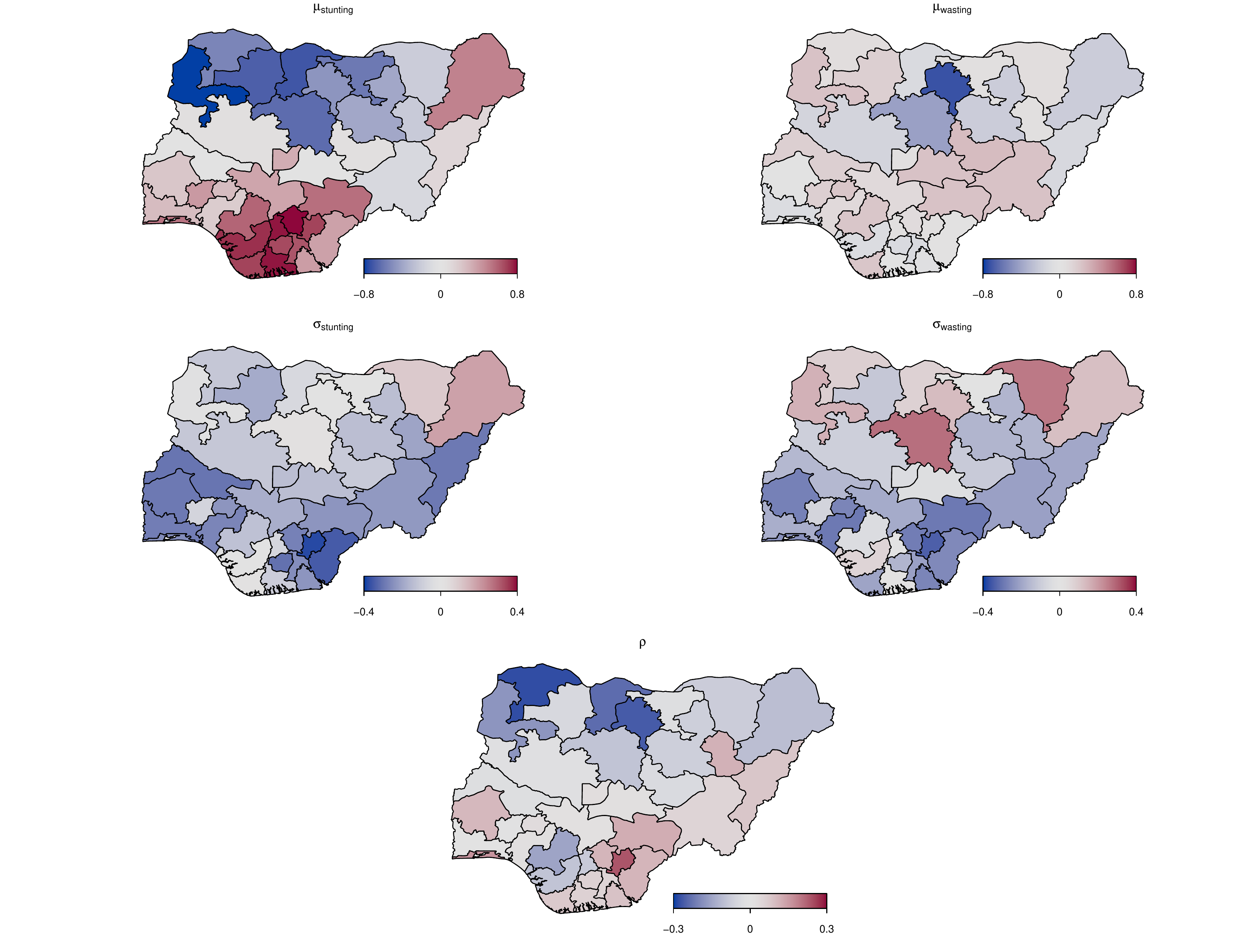}
    \caption{Spatial structure of \textit{stunting} and \textit{wasting} in Nigeria for the distribution parameters of the Gaussian distribution.}\label{Fig:Nigeria2}
\end{figure}

Figure~\ref{Fig:Nigeria1} and \ref{Fig:Nigeria2} show the results for the  non-linear and spatial effects for all parameters. The estimated linear effects are given in Appendix Table~B2. \textit{Stunting} is estimated to be more affected by variables describing children's living situation, particularly \textit{ctwin} (child is a twin) and the birth order (\textit{cbirthorder}). Following our model, with higher birth order, the \textit{stunting} score decreases, with negative values indicating that the children's growth is below the expected growth of a child with normal nutrition. For wasting, \textit{ctwin} had the largest effect, displaying also an increased risk for acute undernutrition. These results are in line with those of \citet{KleCarKneLanWag2021}.

\begin{comment}
\begin{table}[h]
\centering
\begin{tabular}{lrrrrr}
  \toprule
  Covariates & $\mu_{\text{stunting}}$ & $\mu_{\text{wasting}}$ & $\sigma_{\text{stunting}}$ & $\sigma_{\text{wasting}}$ &  \multicolumn{1}{c}{$\rho$} \\  \midrule
  Intercept & -1.2667 & -0.7516 & 0.6301 & 0.2839 & -0.1728 \\ 
  bicycle & -0.0050 & -0.0185 & - & -0.0145 & 0.0319 \\ 
  car & 0.1379 & 0.0236 & - & -0.0374 & - \\ 
  cbirthorder2 & 0.0005 & 0.0427 & 0.0081 & 0.0380 & 0.0165 \\ 
  cbirthorder3 & -0.1090 & -0.0068 & 0.0089 & 0.0419 & 0.0034 \\ 
  cbirthorder4 & -0.1575 & -0.0497 & 0.0419 & - & 0.0095 \\ 
  cbirthorder5 & -0.1368 & -0.0521 & 0.0326 & 0.0019 & 0.0020 \\ 
  cbirthorder6 & -0.2185 & -0.0179 & 0.0363 & -0.0191 & 0.0269 \\ 
  cbirthorder7 & -0.2448 & -0.0319 & 0.0123 & -0.0158 & 0.0376 \\ 
  cbirthorder8 & -0.3185 & -0.0115 & 0.0255 & 0.0275 & 0.0282 \\ 
  csex & 0.1564 & 0.0388 & 0.0117 & 0.0008 & -0.0070 \\ 
  ctwin & -0.3672 & -0.1959 & 0.0436 & -0.0459 & - \\ 
  electricity & 0.0540 & - & 0.0050 & 0.0087 & - \\ 
  motorcycle & 0.0016 & 0.0070 & - & -0.0176 & -0.0104 \\ 
  mresidence & -0.0215 & 0.0624 & - & - & - \\ 
  munemployed & - & - & -0.0034 & -0.0340 & - \\ 
  radio & 0.0197 & - & -0.0060 & 0.0114 & - \\ 
  refrigerator & 0.1247 & - & - & - & 0.0763 \\ 
  television & 0.0820 & -0.0026 & -0.0401 & -0.0382 & 0.0016 \\ 
  \bottomrule
\end{tabular}
\caption{Results of the linear effects for \textit{stunting} and \textit{wasting} of the bivariate Gaussian regression model for the Nigeria data.}\label{Tab:Nigeria}
\end{table}
\end{comment}

Furthermore, \textit{stunting} and \textit{wasting} were both influenced by \textit{cage} and \textit{mbmi} as well.
Following our model, for \textit{mbmi}, a higher BMI of the mother indicates a higher acute and chronic undernutrition. For \textit{cage}, \textit{stunting} and \textit{wasting} is estimated to decrease (i.e.~risk increases) up to around 20 months. After 20 months, the risk for \textit{wasting} is estimated to decrease again while remaining similar for \textit{stunting}. 

The scale parameter for \textit{wasting}, for example, indicates a higher variability for the age of children up to around 25 months. For children older than 25 months, the variability decreases slightly, whereby we observed a greater variability for stunting between 20 and 40 months.
The correlation is negative for children younger than 20 months and is approximately zero after a small positive correlation between 20 and 50 months. This finding indicates an interaction between \textit{stunting} and \textit{wasting} depending on the child's age, which is non-linear and stronger for younger children. Thus, children with a greater height in the first years of life have a lower weight for height and vice versa. The other covariates have only a minor estimated effect on the correlation parameter. These results are consistent with previous findings~\citep{KleKneKlaLan2015,KleCarKneLanWag2021}, which also holds for the spatial effects. 
The regional effect was selected to be informative for all distribution parameters.
The effect of chronic undernutrition, for example, showed a lower risk of stunted growth in regions in southern Nigeria due to a positive effect. These regions also have a lower variability of chronic undernutrition compared to the average regions in the center of the county. This means that in this part of Nigeria the score for stunting is estimated to be on average lower and its variability is also smaller. By contrast, the regions in the north are estimated to have a higher risk for \textit{stunting}. In terms of the correlation, some regions in the north are estimated to have a negative effect, while other regions in the south are estimated to show a slight positive effect on the correlation. A positive effect suggests that these regions have a problem of acute undernutrition as well as chronic undernutrition.

Overall, chronic undernutrition (\textit{stunting}) is mostly affected by the living conditions of the children, for example, the birth order. Whereby, \textit{stunting} and \textit{wasting} were both influenced by the mother's BMI and particularly by the child's age. Additional effects of the covariates on the scale and correlation parameters also suggested greater uncertainty for younger children for acute undernutrition through a positive effect on the standard deviation, with variability decreasing with age. Furthermore, we observed a stronger negative correlation between \textit{stunting} and \textit{wasting} in younger children, i.e., as stunting increases, wasting is expected to be lower. This means that children with a greater height tend to suffer from a lower weight for height at a younger age.

% Negative values for the mean value of stunting for regions in the northern region in the top left (namely Sokoto, Kebbi, Zamfara and Katsina). Positive values for the bottom regions borders the atlantic ocean (Mid-western  and eastern region, e.g., Rivers, Enugu, Anambra, Delta)

\section{Discussion}\label{Discussion}

We developed statistical boosting for modeling distributional regression with multivariate outcomes. Motivated by our biomedical applications, we considered three important multivariate parametric distributions: the bivariate Bernoulli, Poisson and Gaussian distributions. As special merits over classical maximum likelihood or Bayesian approaches to multivariate GAMLSS, our boosting framework can directly be used for high-dimensional data problems ($p\gg n$), while allowing for a data-driven variable selection mechanism that allows for sparse models for all parameters of a multivariate distribution. %Our model is also more flexible than e.g.~the bivariate Poisson model proposed by~\cite{KarlisNtzoufras2005} in terms of covariate effects, which in our case can go beyond linearity by allowing for non-linear and spatial covariate structures.

In simulation studies, we have illustrated that the proposed boosting approach is able to identify the correct predictors in different data situations, including low- and high-dimensional settings and incorporating different effect types such as spatial effects. A comparison with the boosted univariate models showed that the bivariate models yielded more accurate estimates for the true structure of the effects. %These results are not surprising, as the bivariate models can also take the association between the two outcomes into account. 

The wide applicability of our approach is illustrated on three different biomedical data sets, where we extend previous studies and also confirm findings from the literature. 
Applying our approach to examine jointly the genetic predisposition for chronic ischemic heart disease and high cholesterol not only provides information on the dependency of these phenotypes on the genetic variants, but also allows to identify the variants that affect the association between both phenotypes. This is in strong contrast to classical methods to estimate e.g., polygenic risk scores via accumulating effects from univariate linear models with single variants as predictor variables~\citep{PRS}. Our approach does not only incorporate multivariable predictor models, but also considers multivariate outcomes and hence allows to assess also the genetic predisposition for the association between several phenotypes, such as heart disease and high cholesterol. To the best of our knowledge, this is the first time multivariate distributional regression was adapted to model the joint genetic liability for multiple phenotypes.   

In examining possible effects of patients characteristics on demand for health care, we found that age and income are relevant predictors, but also that gender affected the association between the number of doctor consultations and prescribed medications, with a stronger association found for women~\citep[cf.][]{KarlisNtzoufras2005}. 

In the third application analyzing the risk of undernutrition in Nigeria, an association was found between chronic undernutrition and the child's living condition. In addition, the age of the child had a relevant influence on all distribution parameters related to chronic and acute undernutrition; furthermore, the regional effect was selected not only for the margins but also for the scale and correlation parameters.
%In summary, the results for the three biomedical data applications illustrate the benefits of fitting bivariate distributional regression models in combination with boosting to analyze diverse research questions.

% Limitation
A limitation regarding the considered distributions in our approach is the restriction of the Poisson distribution to positive dependency between the two responses. A possible solution for this restriction in future research could be the use of alternative parametrizations~\citep{Lakshminarayana1999}, which also allow for modeling negative correlations; however, these have the disadvantage that the interpretation of effects on these parameters becomes much more difficult. 
A limitation of our algorithm is the relatively high selection rates for variables with only minor importance, which occurs particularly in low-dimensional settings. In this context, \cite{Deselection} have recently proposed an approach to deselect predictors with negligible impact to obtain sparser models with statistical boosting. We want to investigate the incorporation of this proposal in the context of multivariate GAMLSS in the future. Moreover, as the number of distribution parameters and the complexity of the model increases (e.g., due to many non-linear effects), the algorithm becomes computationally more intensive. %, particularly with respect to the tuning procedure for the stopping iteration using resampling techniques such as cross-validation. 
To address this problem, also alternative approaches for early stopping could be considered. A promising approach in the future which has been developed for univariate location models is probing, where randomly shuffled versions of the original observed variables (probes) are added to the data set and the algorithm stopped when the first probe is selected~\citep{probing}. %Further research is warranted to investigate how this approach can be incorporated into more complex models with multiple additive predictors where the variables are selected simultaneously.

% Future research
Last, our focus has been on bivariate distributional regression models, but we will consider extending the models to higher dimensional responses in future research. From an algorithmic perspective, the extension should be straight-forward as it only adds more distributional parameters in our proposed framework. However, not only the construction of appropriate response distributions but also the interpretation of the effect estimates becomes more challenging. For example, for the multivariate Gaussian distribution, the main challenge is the parameterization of the covariance matrix and a promising route here could be based on a (modified) Cholesky decomposition~\citep{Pourahmadi2011}.
Similar, the extension of the bivariate Poisson distribution to higher dimensions has some difficulties due to the complicated form of the joint probability function. The most common extension would force all the pairs of variables to have the same covariance~\citep{Karlis2003multi}, whereby \cite{Karlis2005MultivariatePR} already discussed a model with a two-way covariance term that allows for different covariances between the variables. %However, also these extensions can still only represent positive correlations between the outcomes. 

% For the Bernoulli distribution it exist already a trivariate approach by Islam (2018).  https://www.tandfonline.com/doi/pdf/10.1080/25742558.2018.1472519 - trivariate Bernoulli
% \cite{MarshallOlkin1985} ??? Many distributions can naturally derived from the bivariate Bernoulli distribution

%Our boosting approach for modeling bivariate distributional regression models provides a flexible framework to investigate complex biomedical research questions, where multiple related outcomes of interest are modeled jointly. In addition, this approach combines the main properties of GAMLSS and of statistical boosting for multivariate responses. For all distribution parameters, the most relevant predictors can be identified simultaneously from a potentially high-dimensional set of candidate variables. Apart from that, modeling dependencies between outcomes in particular enables deeper insights and a more flexible model representation. 

\subsection*{Supplementary Materials}
A supplement contains further information on the simulation and biomedical applications.

\subsection*{Acknowledgement}
The work on this article was supported by the Deutsche Forschungsgemeinschaft (DFG, grant number 428239776, KL3037/2-1, MA7304/1-1).

\spacingset{1.1}
\bibliographystyle{apalike}
\bibliography{Literatur}
\renewcommand\bibfont{\footnotesize}

\appendix

\setcounter{figure}{0}

\makeatletter 
\renewcommand{\thefigure}{A\@arabic\c@figure}
\makeatother

\makeatletter 
\renewcommand{\thetable}{A\@arabic\c@table}
\makeatother

\section*{\centering Supplementary Materials}
\section{Further simulation results}
\subsection{Bivariate Bernoulli distribution}
\subsubsection{Low-dimensional setting}
For the simulation of a low-dimensional setting of the bivariate logit model, we considered $n = 1000$ observations and $p = 10$ covariates for each of the three parameters. For data generation, the \textsc{R} package \textbf{VGAM}~\citep{VGAM} was used, whereby the parameters $p_1, p_2$ and $\psi$ were simulated with the following linear predictors
\begin{align*}
    \logit(p_1)  &= \eta_{\mu_1} = X_1 + 1.5 X_2 - X_3 + 1.5 X_4, \\
    \logit(p_2)  &= \eta_{\mu_2} = 2X_1 - X_2 + 1.5 X_3, \\
    \log(\psi)   &= \eta_{\psi} =  - 1.5 + 1 X_5 + 1.5 X_6.
\end{align*}
Overall, only the first six covariates out of the $p = 10$ had a relevant effect on any of the distribution parameters (four for $p_1$, three for $p_2$ and two for $\psi$). The covariates were simulated from a multivariate normal distribution $N(\nullvec,\mSigma)$ with a Toeplitz covariance structure $\Sigma_{ij} = \rho^{|i-j|}$ for $1\leq i,j\leq p$, where $\rho =0.5$ is the correlation between consecutive variables $X_j$ and $X_{j+1}$.  
The covariates were incorporated in the boosting approach by using simple linear models as base-learners.
As measures for the predictive performance, the area under the curve (AUC), the Brier score, the negative log-likelihood and energy score were considered. Note that the AUC and Brier score do not account for the dependence between the two outcomes and are calculated separately for both outcomes.

\begin{table}[h]
\centering
\begin{tabular}{lcccc}
   \toprule
        \multirow{2}{*}{Parameter}  & \multicolumn{2}{c}{Univariate} & \multicolumn{2}{c}{Bivariate}  \\
        & inf & non-inf & inf & non-inf  \\ \midrule
        $\mu_1$ & 100\% & 86.83\% & 99.75\% & 32.17\%  \\
        $\mu_2$ & 100\% & 92.57\% & 100\% & 42.14\%  \\
        $\psi$ & - & - & 75.50\% & 19.88\% \\
 \bottomrule
\end{tabular}
\caption{Resulting selection rates of the low-dimensional setting of the bivariate Bernoulli distribution; the average values from the 100 simulation runs are reported for the univariate and bivariate model for the informative and non-informative variables.}
\end{table}

\begin{table}[h]
    \centering
    \begin{tabular}{lcc}
    \toprule
         & Univariate & Bivariate \\ \midrule
        AUC ($Y_1$) & 0.89 (0.01) & 0.88 (0.01)  \\ 
        AUC ($Y_2$) & 0.86 (0.01) & 0.86 (0.01) \\ 
        Brier score ($Y_1$) & 0.14 (0.01) & 0.14 (0.01) \\ 
        Brier score ($Y_2$) & 0.15 (0.01) & 0.15 (0.01) \\ 
        Energy score & 0.22 (0.23) &  0.27 (0.01)\\ 
        Negative log-likelihood & 882.03 (24.05) & 874.97 (26.80)  \\ \bottomrule
    \end{tabular}
    \caption{Resulting predictive performance on independent test data for the low-dimensional linear setting of the bivariate Bernoulli distribution; mean (sd) values from 100 simulation runs are reported for the univariate and bivariate models.}
\end{table}

\begin{figure}[t!]\centering
    \includegraphics[width = \textwidth]{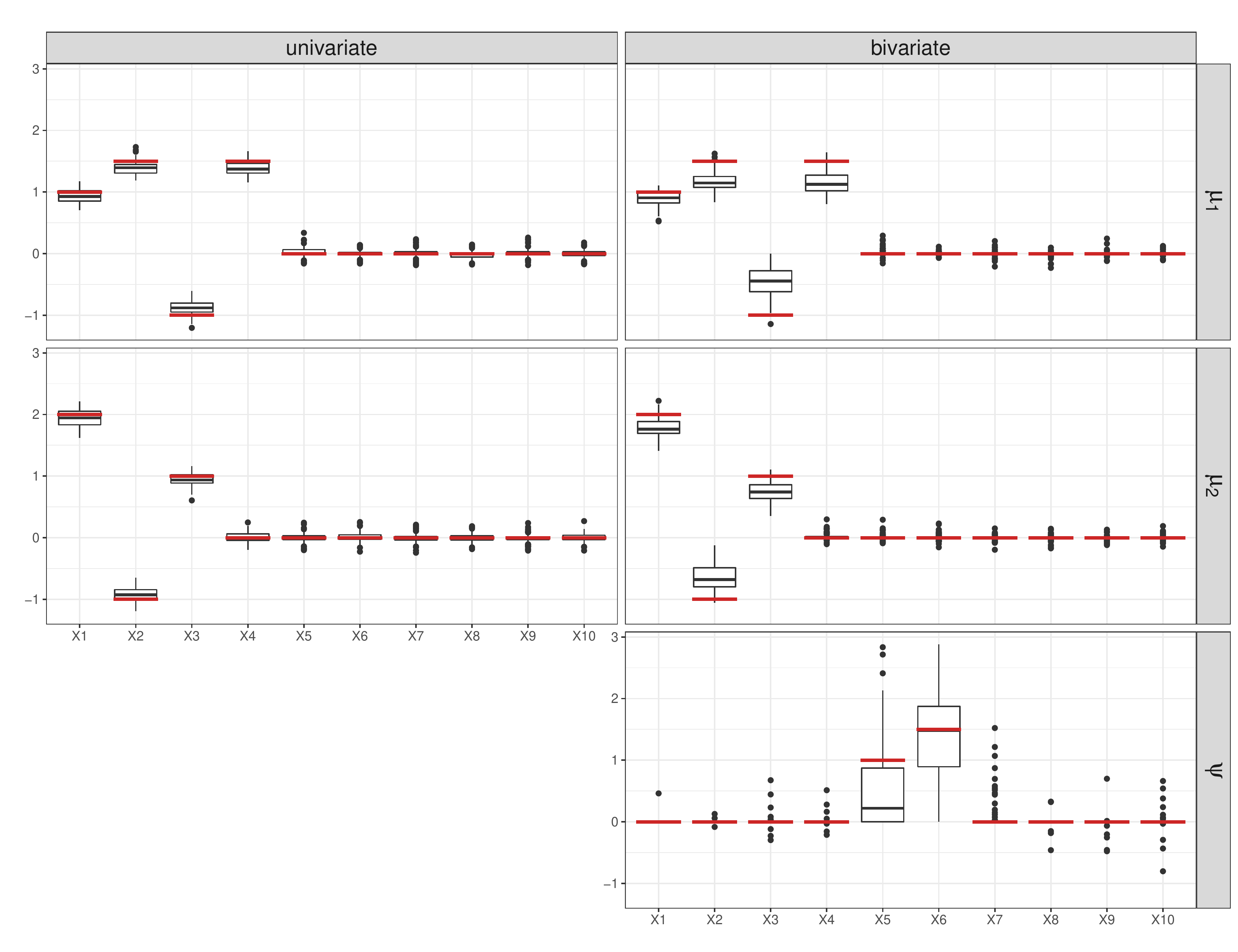} 
    \caption{Results for the estimated linear effects of the univariate (left) and bivariate Bernoulli (right) model of the ten covariates $X_1, \dots, X_{10}$ for the low-dimensional setting from 100 simulation runs. The red horizontal lines correspond to the true values.}
\end{figure}

\newpage
\subsubsection{High-dimensional setting}

\begin{table}[h]
\centering
\begin{tabular}{lcccc}
   \toprule
        \multirow{2}{*}{Parameter}  & \multicolumn{2}{c}{Univariate} & \multicolumn{2}{c}{Bivariate}  \\
        & inf  & non-inf & inf  & non-inf \\ \midrule
        $\mu_1$ &  100\% & 5.18\% & 97.75\% & 1.26\% \\
        $\mu_2$ &  100\% & 5.93\% & 100\% & 1.57\% \\
        $\psi$ &  -  & - & 59\% & 0.17\%\\
 \bottomrule
\end{tabular}
\caption{Resulting selection rates of the high-dimensional setting of the bivariate Bernoulli distribution; the average values from the 100 simulation runs are reported for the univariate and bivariate model for the informative and non-informative variables.}
\end{table}

\clearpage
\newpage

\subsection{Bivariate Poisson distribution}
\subsubsection{Low-dimensional setting}
\begin{table}[ht!]
\centering
\resizebox{\textwidth}{!}{%
\begin{tabular}{lcccc|ccccc}
   \toprule
       &\multicolumn{4}{c|}{Linear model} &  \multicolumn{4}{c}{Non-linear model}\\
        \multirow{2}{*}{Parameter} & \multicolumn{2}{c}{Univariate} & \multicolumn{2}{c|}{Bivariate}  &  \multicolumn{2}{c}{Bnivariate} & \multicolumn{2}{c}{Bivariate} \\
        & inf & non-inf & inf & non-inf & inf  & non-inf & inf  & non-inf \\ \midrule
        $\lambda_1$ & 98.67\% & 75.29\% & 98.67\% & 54.43\% & 100\% & 46.56\% & 100\% & 77.56\% \\
        $\lambda_2$ & 100\% & 73.17\% & 100\% & 57.67\% & 100\% & 45.22\% & 100\% &  84.11\% \\
        $\lambda_3$ & - & - & 95.67\% & 49.43\% & -  & - & 100\% & 53.78\%\\
 \bottomrule
\end{tabular}}
\caption{Resulting selection rates of the low-dimensional linear and non-linear setting of the bivariate Poisson regression; the average values from the 100 simulation runs are reported for the univariate and bivariate model for the informative and non-informative variables.}
\end{table}

\subsubsection{High-dimensional setting}

We investigated in the linear and non-linear settings with $p=1000$ covariates and $n= 1000$ observations for each distribution parameter.
For the linear setting, the  underlying true predictors were specified as
\begin{align*}\label{PoisLin}
    \log(\lambda_1)  &= \eta_{\lambda_1} =  -X_1 + 0.5X_2 + 1.5 X_3,  \\
    \log(\lambda_2)  &= \eta_{\lambda_2} = 2X_1 - X_3 + 1.5 X_4 + X_5,  \\
    \log(\lambda_3)  &= \eta_{\lambda_3} =0.5 X_5 + X_6 - 0.5 X_7, 
\end{align*}
where the covariates followed a multivariate normal distribution $N(\nullvec,\mSigma)$ with Toeplitz covariance structure and correlation coefficient $\rho =0.5$. Thus, the first seven covariates were informative for any of the distribution parameter (three for $\lambda_1$ and $\lambda_3$, four for $\lambda_2$). For this setting, simple linear models were incorporated as base-learners. 
For the non-linear setting, the true additive predictors were given by
\begin{align*}
    \log(\lambda_1)  &= \eta_{\lambda_1} = \sqrt{X_1}X_1, \\
    \log(\lambda_2)  &= \eta_{\lambda_2} = \cos(2X_2), \\
    \log(\lambda_3)  &= \eta_{\lambda_3} = \sin{X_3}, 
\end{align*}
where the covariates were independently simulated from the uniform distribution~$U(0,1)$ and only one covariate was informative for each of the distribution parameters. As base-learners, we chose P-splines (20 equidistant knots with a second-order difference penalty and four degrees of freedom).

\begin{table}[ht!]
\centering
\resizebox{\textwidth}{!}{%
\begin{tabular}{lcccc|ccccc}
   \toprule
      \multirow{2}{*}{Parameter} &\multicolumn{4}{c|}{Linear model} &  \multicolumn{4}{c}{Non-linear model}\\
          & \multicolumn{2}{c}{Univariate} & \multicolumn{2}{c|}{Bivariate}  &  \multicolumn{2}{c}{Univariate} & \multicolumn{2}{c}{Bivariate} \\
        & inf & non-inf & inf & non-inf & inf  & non-inf & inf  & non-inf \\ \midrule
        $\lambda_1$ & 70.67\% & 0.83\% & 70.33\% & 1.48\% & 100\% & 0.54\% & 100\% & 0.90\% \\
        $\lambda_2$ & 96.00\% & 1.03\% & 96.00\% & 1.93\% & 97\% & 0.41\% & 100\% &  2.26\% \\
        $\lambda_3$ & - & - & 62.33\% & 0.79\% & -  & - & 75.00\% & 0.01\%\\
 \bottomrule
\end{tabular}}
\caption{Resulting selection rates of the high-dimensional linear and non-linear setting of the bivariate Poisson regression; the average values from the 100 simulation runs are reported for the univariate and bivariate model for the informative and non-informative variables.}
\end{table}

\begin{figure}[h!]\centering
    \includegraphics[width = \textwidth]{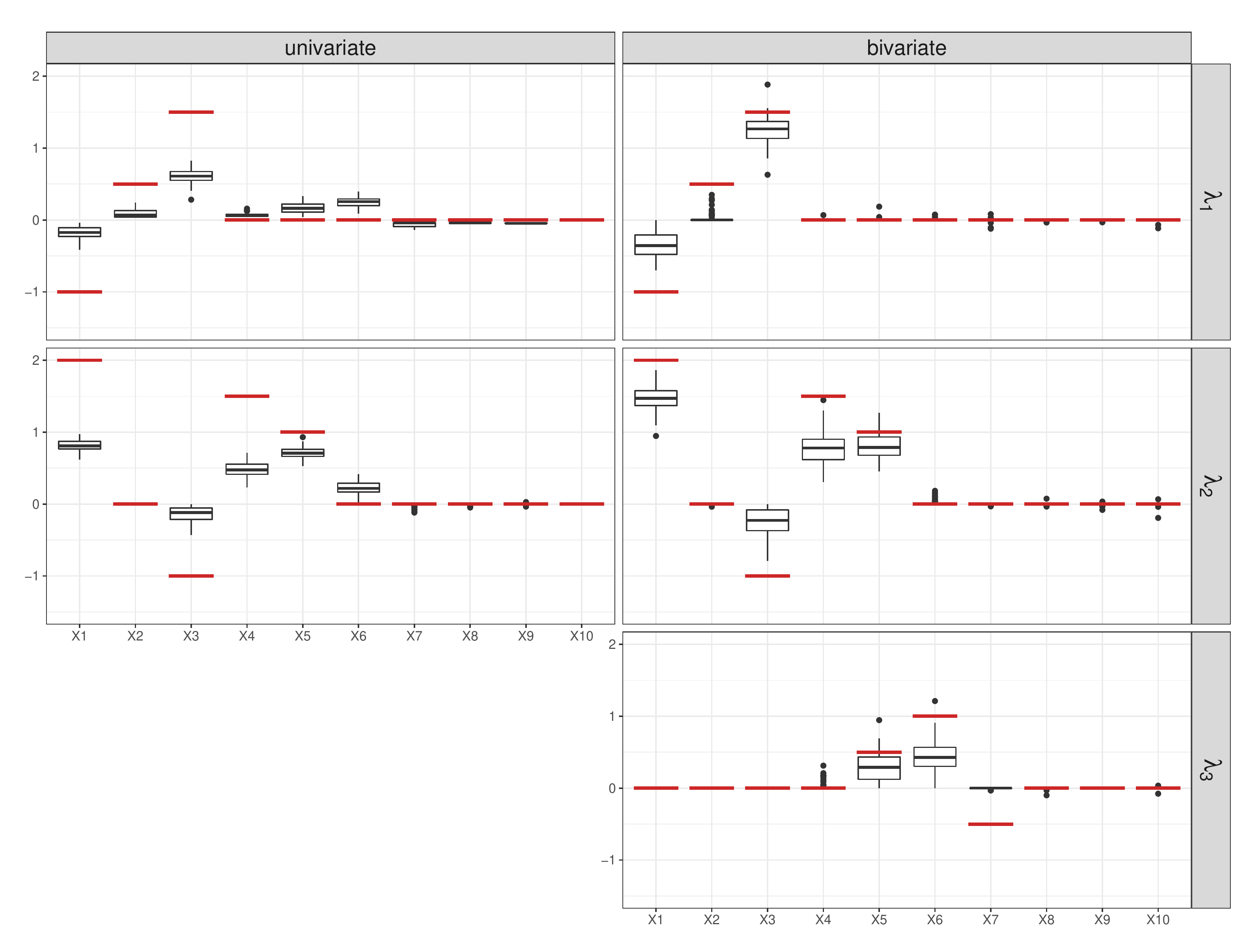} 
    \caption{Results for the estimated linear effects of the univariate (left) and bivariate Poisson model (right) of the first ten covariates $X_1, \dots, X_{10}$ for the high-dimensional setting from 100 simulation runs. The horizontal lines correspond to the true values.}
\end{figure}

\begin{figure}[h!]\centering
    \includegraphics[scale = 0.4]{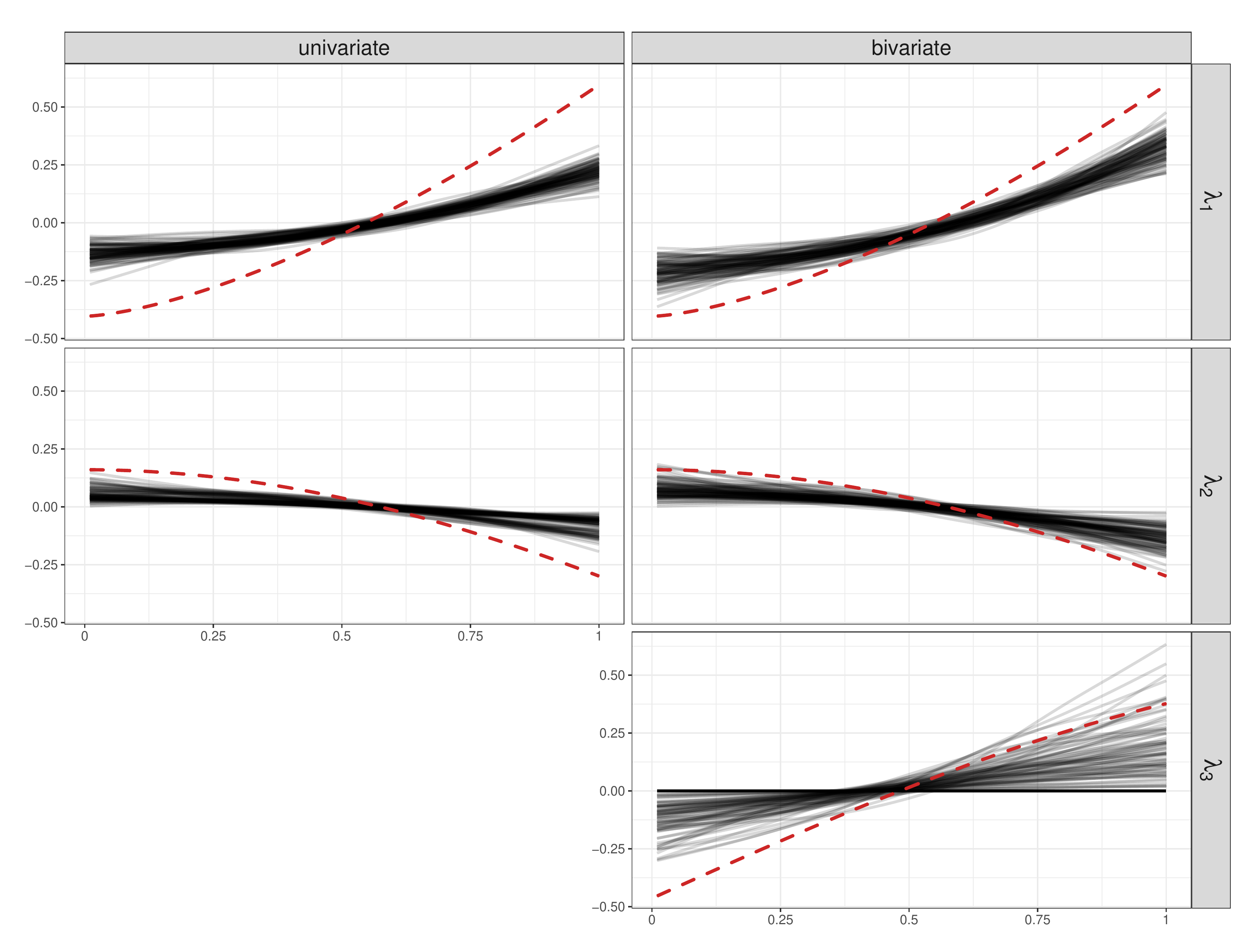} 
    \caption{Results for the estimated non-linear effects of the univariate (left) and the bivariate Poisson model (right) of the first ten covariates $X_1, \dots, X_{10}$ for the high-dimensional setting from 100 simulation runs. The red dotted lines correspond to the true effects.}
\end{figure}

\begin{table}[!h]
    \centering
    \resizebox{\textwidth}{!}{%
    \begin{tabular}{l|cc|ccc}
    \toprule
        &\multicolumn{2}{c|}{Linear model} &  \multicolumn{2}{c}{Non-linear model}\\
        &  Univariate & Bivariate & Univariate & Bivariate \\ \midrule
        MSEP ($Y_1$) & 2.65 (0.16) & 3.94 (0.25) & 4.58 (0.26) & 7.12 (0.39) \\
        MSEP ($Y_2$) & 3.06 (0.26) & 4.24 (0.34) & 5.47 (0.35) & 8.03 (0.51) \\ 
        Energy score & 1.40 (0.04) & 1.40 (0.03) & 1.94 (0.05) & 1.94 (0.05)\\ 
        Negative log-likelihood  & 3630.27 (49.83) & 3454.63 (39.48) & 4422.52 (58.82) & 4250.49 (47.49) \\\bottomrule
    \end{tabular}}
    \caption{Resulting predictive performance on independent test data for the linear and non-linear settings of the bivariate Poisson regression for the high-dimensional setting; mean (sd) values from 100 simulation runs are reported for the univariate and bivariate models.\label{Table:Pois}}
\end{table}

\clearpage
\newpage

\subsection{Bivariate Gaussian distribution}
\subsubsection{Low-dimensional setting}

\begin{figure}[h!]\centering
    \includegraphics[width = \textwidth]{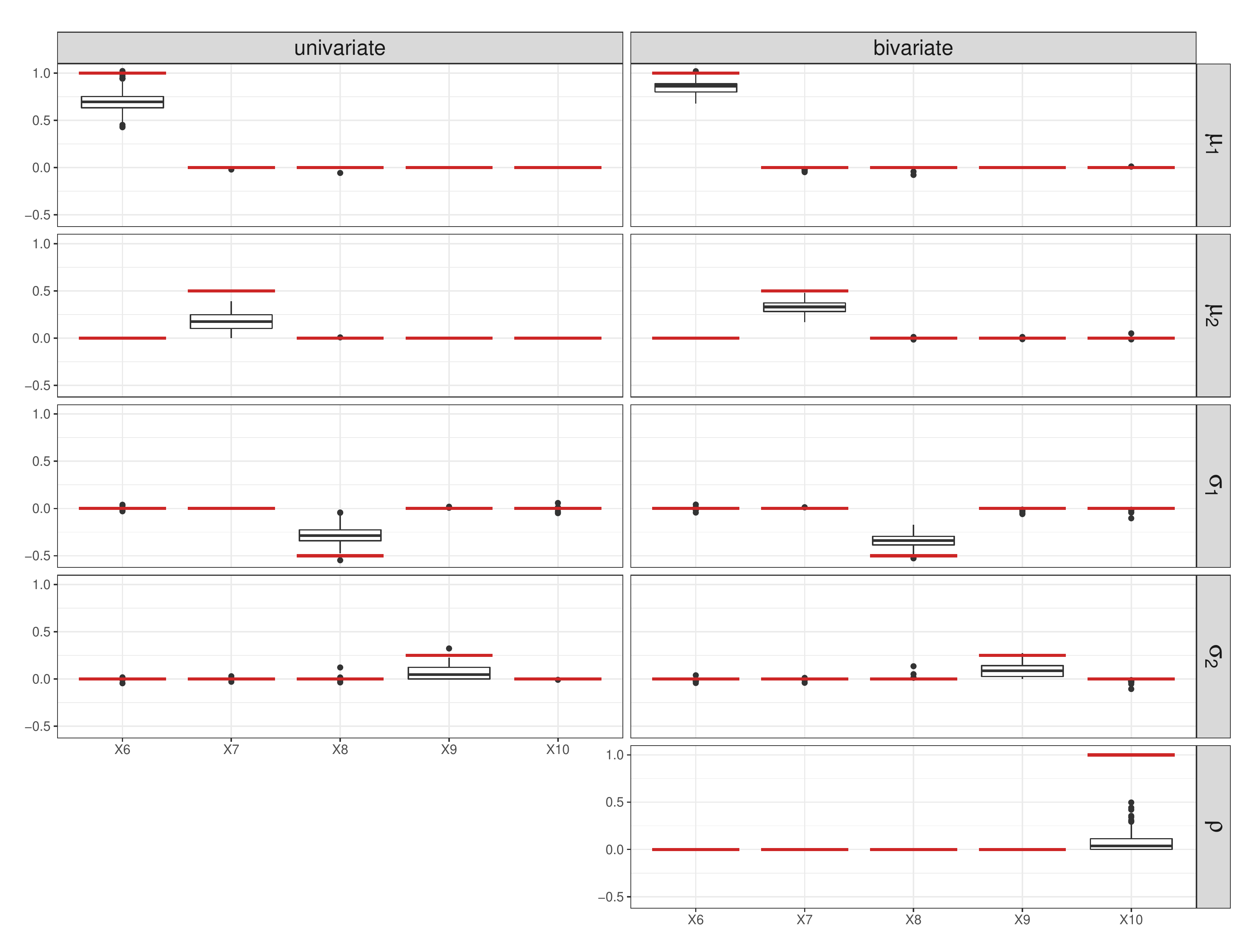} 
    \caption{Results for the estimated linear effects of the univariate (left) and bivariate Gaussian regression model (right) for the low-dimensional setting from 100 simulation runs. The horizontal lines correspond to the true values.}
\end{figure}

\begin{figure}[h!]\centering
    \includegraphics[width = \textwidth]{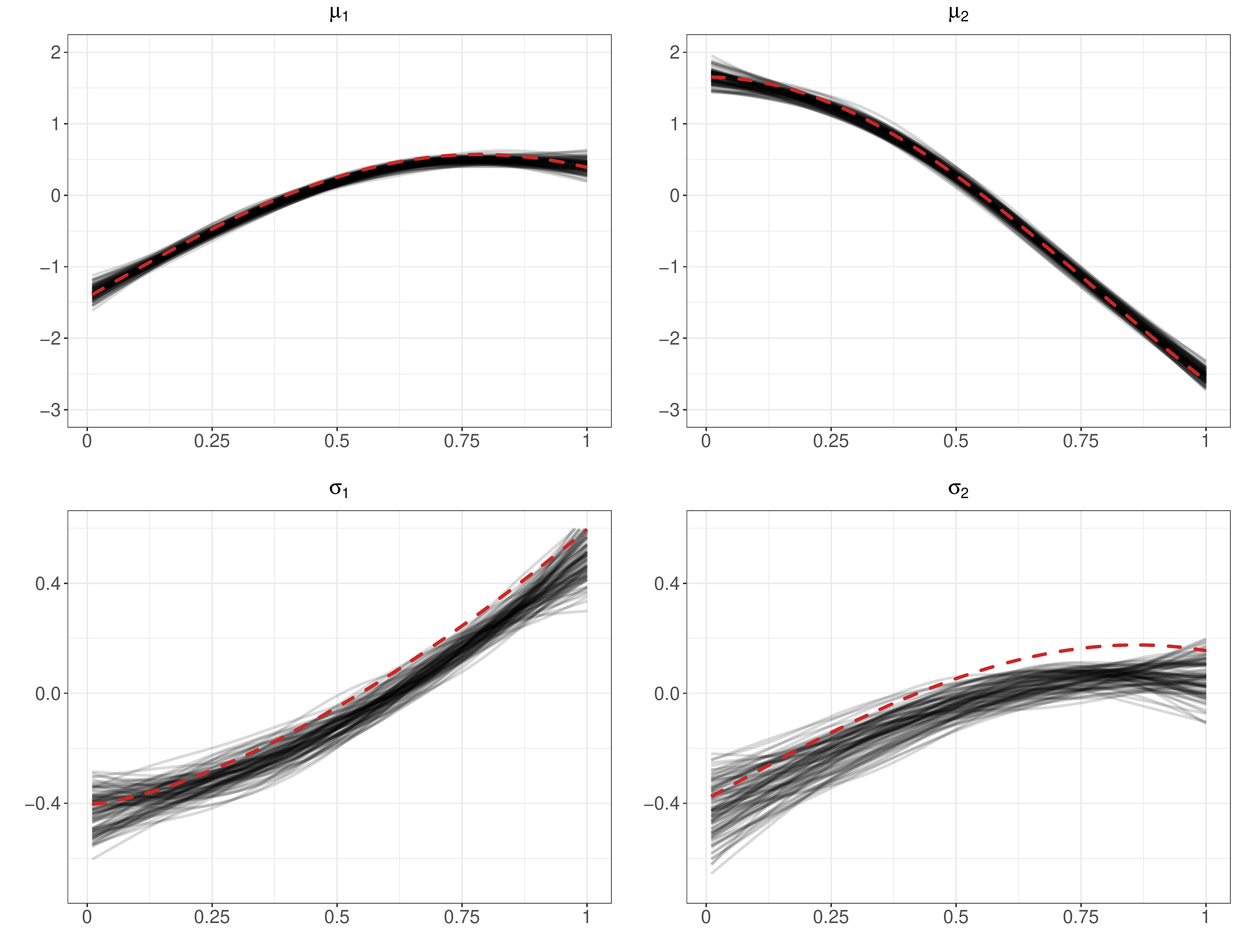}
    \caption{Results for the estimated non-linear effects of the univariate Gaussian regression models for the low-dimensional setting from 100 simulation runs. The red dotted lines correspond to the true effects.}
\end{figure}

\begin{table}[ht]
\centering
\begin{tabular}{lcccc}
   \toprule
        \multirow{2}{*}{Parameter}  & \multicolumn{2}{c}{Univariate} & \multicolumn{2}{c}{Bivariate}    \\
        & inf & non-inf & inf & non-inf \\ \midrule
        $\mu_1$ & 100\% & 50.00\% & 100\% & 57.29\%  \\
        $\mu_2$ & 100\% & 48.14\% & 100\% & 50.86\% \\
        $\sigma_1$ & 100\% & 59.29\% & 100\% & 61.57\% \\
        $\sigma_2$ & 100\% & 57.43\% & 100\% &65.14\% \\
        $\rho$ & - & - & 98.5\% & 20.57\% \\
 \bottomrule
\end{tabular}
\caption{Resulting selection rates of the low-dimensional linear and non-linear setting of the bivariate Gaussian regression; the average values from the 100 simulation runs are reported for the univariate and bivariate model for the informative and non-informative variables.}
\end{table}

% Spatial Effect
\begin{figure}[h!]\centering
    \includegraphics[width = \textwidth]{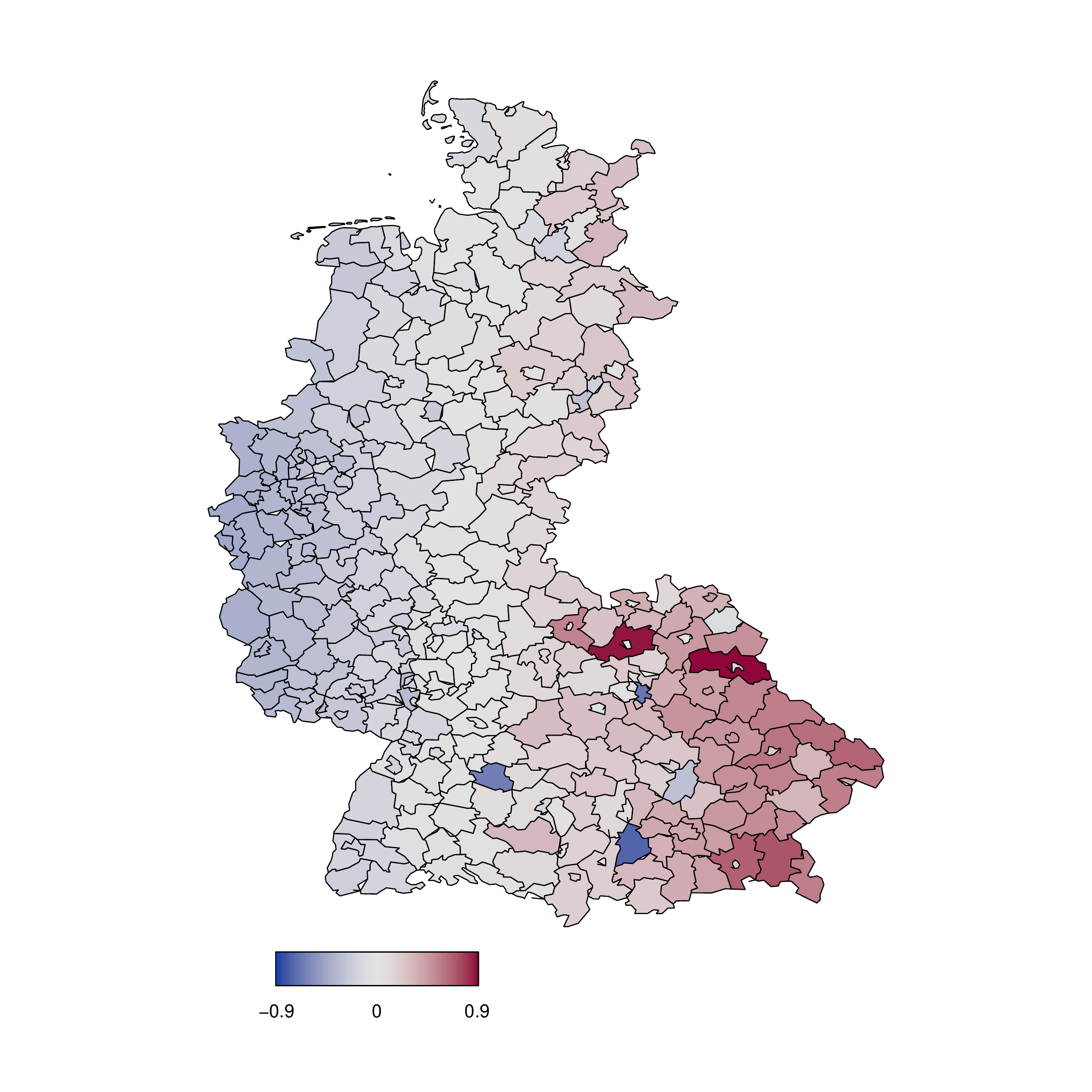} 
    \caption{Map of the true structure of the spatial effect of West Germany.}
\end{figure}

\begin{figure}[h!]\centering
    \includegraphics[width = \textwidth]{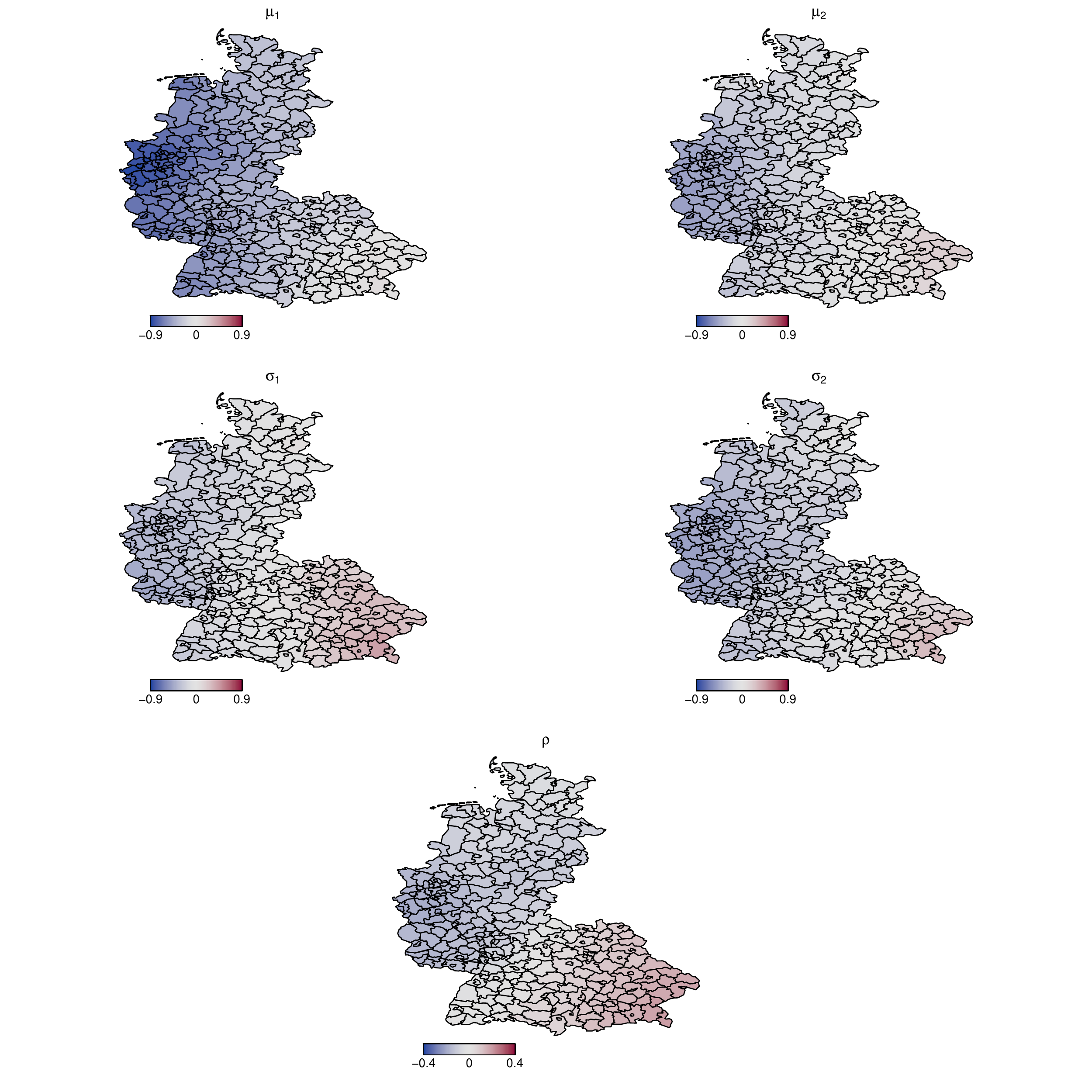} 
    \caption{The estimated spatial effects for one simulation run of the low-dimensional setting of the bivariate Gaussian regression model.}
    \end{figure}

\clearpage
\newpage

\subsection{High-dimensional setting}
For the simulation of a low-dimensional setting of a bivariate Gaussian distributed outcome, we considered a setting with linear, non-linear and spatial effects with $p=1000$ covariates and $n=1000$ observations with the following true predictors
\begin{small}
\begin{align*}
    \mu_1  = \eta_{\mu_1} = \sin(2X_1)/0.5 + X_6 + 0.5 X_7 + f_{\text{spat}} \qquad &
    \mu_2  = \eta_{\mu_2} = 2 + 3\cos(2X_2) + 0.5X_7+X_8 + f_{\text{spat}}\\
    \log(\sigma_1) = \eta_{\sigma_1} = \sqrt{X_3}X_3 -0.5X_8 + f_{\text{spat}} \qquad &
    \log(\sigma_2) =\eta_{\sigma_2} = \cos(X_4)X_4 + 0.25X_9 + f_{\text{spat}}\\
    \rho/\sqrt{1-\rho^2} = \eta_{\rho} = \log(X_5^2) &+ X_{10}+ f_{\text{spat}},
\end{align*}
\end{small}\\
where the covariates were independently simulated from the uniform distribution~$U(0,1)$. Each included covariate was informative for one of the distribution parameters; more precisely, for each parameter three covariates, one linear and one non-linear, and additionally the spatial effect. For linear effects we used simple linear models as base-learners and P-splines for the non-linear effects. 
The spatial effects were simulated with $f_{\text{spat}}(s) = \sin(x^{c}_s)\cos(0.5y^{c}_s), s \in {1,\dots,S}$, based on the centroids of the standardized coordinates of the discrete regions in Western Germany with overall $S=327$ regions. The neighborhood structure was modeled by the spatial base-learner using a Markov random field~\citep{Umlauf2019}. 

\begin{table}[!t]
    \centering
    \begin{tabular}{lcc}
    \toprule
         & Univariate & Bivariate \\ \midrule
        MSEP ($Y_1$) & 1.59 (0.10)  & 1.58 (0.10) \\ 
        MSEP ($Y_2$) & 1.38 (0.09) & 1.38 (0.09) \\ 
        Energy score & 1.03 (0.02) & 1.00 (0.02) \\ 
        Negative log-likelihood & 3339.99 (97.13) & 2997.21 (94.28)\\ \bottomrule
    \end{tabular}
    \caption{Resulting predictive performance on independent test data  for the high-dimensional setting of the bivariate Gaussian regression; mean (sd) values from 100 simulation runs are reported for the univariate and bivariate models.\label{Table:Norm}}
\end{table}

\begin{table}[ht]
\centering
\begin{tabular}{lcccc}
   \toprule
        \multirow{2}{*}{Parameter}  & \multicolumn{2}{c}{univariate} & \multicolumn{2}{c}{bivariate}    \\
        & inf & non-inf & inf & non-inf \\ \midrule
        $\mu_1$ & 100\% & 0.68\% & 100\% & 1.07\%  \\
        $\mu_2$ & 100\% & 0.59\% & 100\% & 0.82\% \\
        $\sigma_1$ & 100\% & 1.58\% & 100\% & 1.31\% \\
        $\sigma_2$ & 100\% & 1.77\% & 100\% & 1.35\% \\
        $\rho$ & - & - & 94.5\% & 0.09\% \\
 \bottomrule
\end{tabular}
\caption{Resulting selection rates of the high-dimensional linear and non-linear setting of the bivariate Gaussian regression; the average values from the 100 simulation runs are reported for the univariate and bivariate model for the informative and non-informative variables.}
\end{table}

\begin{figure}[h!]\centering
    \includegraphics[width = \textwidth]{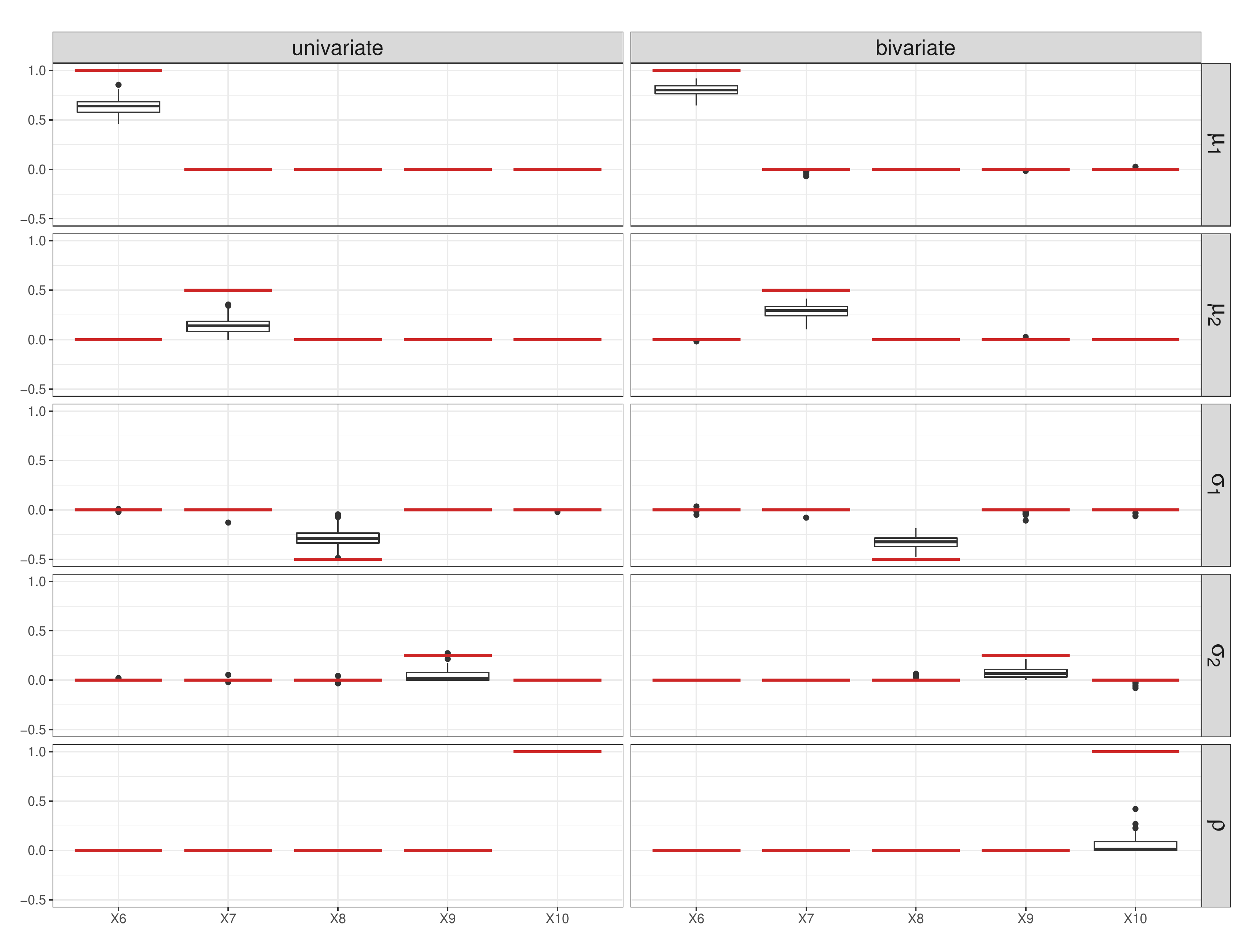}
    \caption{Results for the estimated linear effects of the univariate (left) and bivariate Gaussian regression model (right) for the high-dimensional setting from 100 simulation runs. The horizontal lines correspond to the true values.}
\end{figure}

\begin{figure}[t]\centering
    \includegraphics[width = \textwidth]{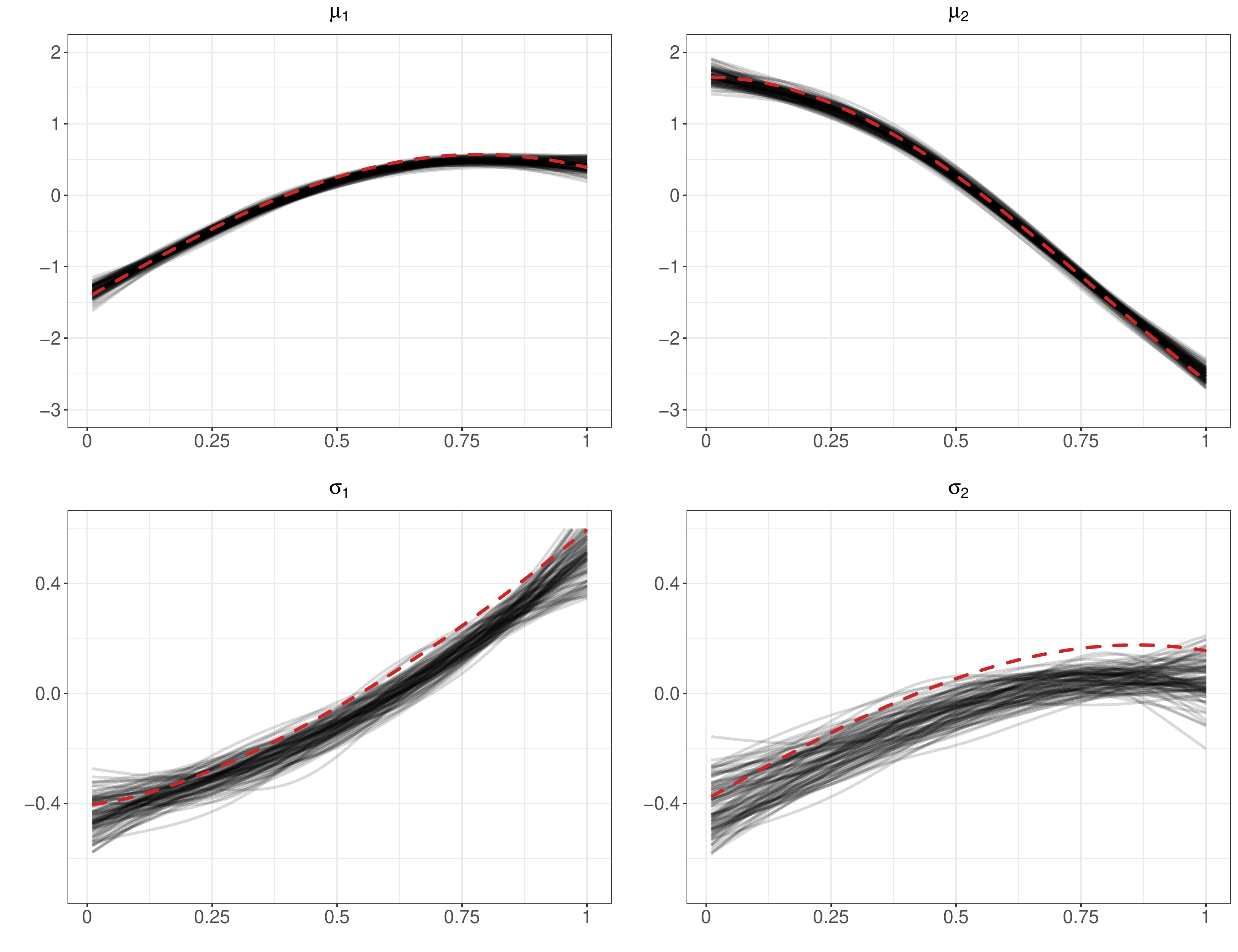}
    \caption{Results for the estimated non-linear effects of the univariate Gaussian regression model for the high-dimensional setting for 100 simulation runs. The red dotted lines correspond to the true effects.}
\end{figure}

\begin{figure}[h!]\centering
    \includegraphics[width = \textwidth]{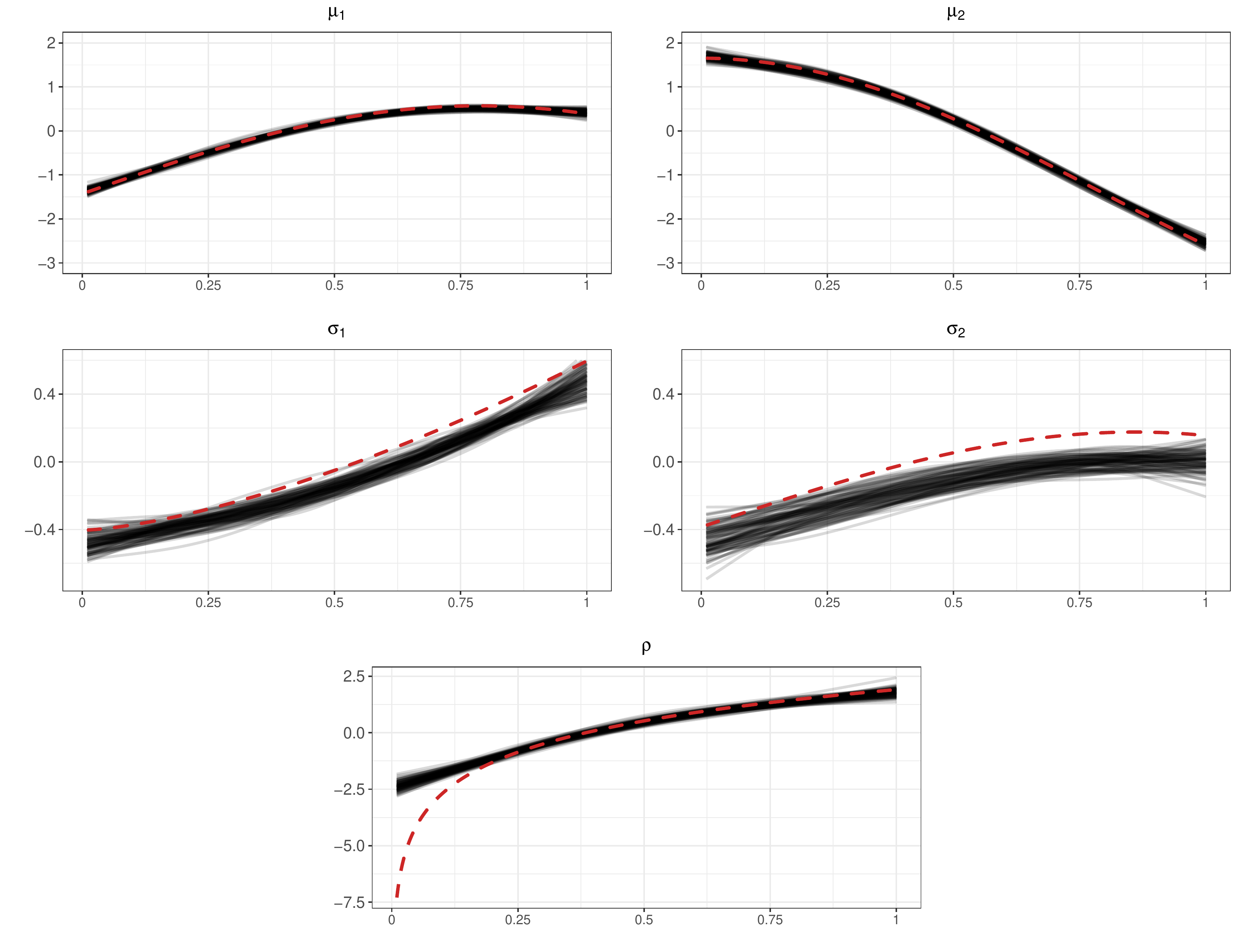}
    \caption{Results for the estimated non-linear effects of the bivariate Gaussian regression model for the high-dimensional setting for 100 simulation runs. The red dotted lines correspond to the true effects.}
\end{figure}

\begin{figure}[t]\centering
    \includegraphics[width = \textwidth]{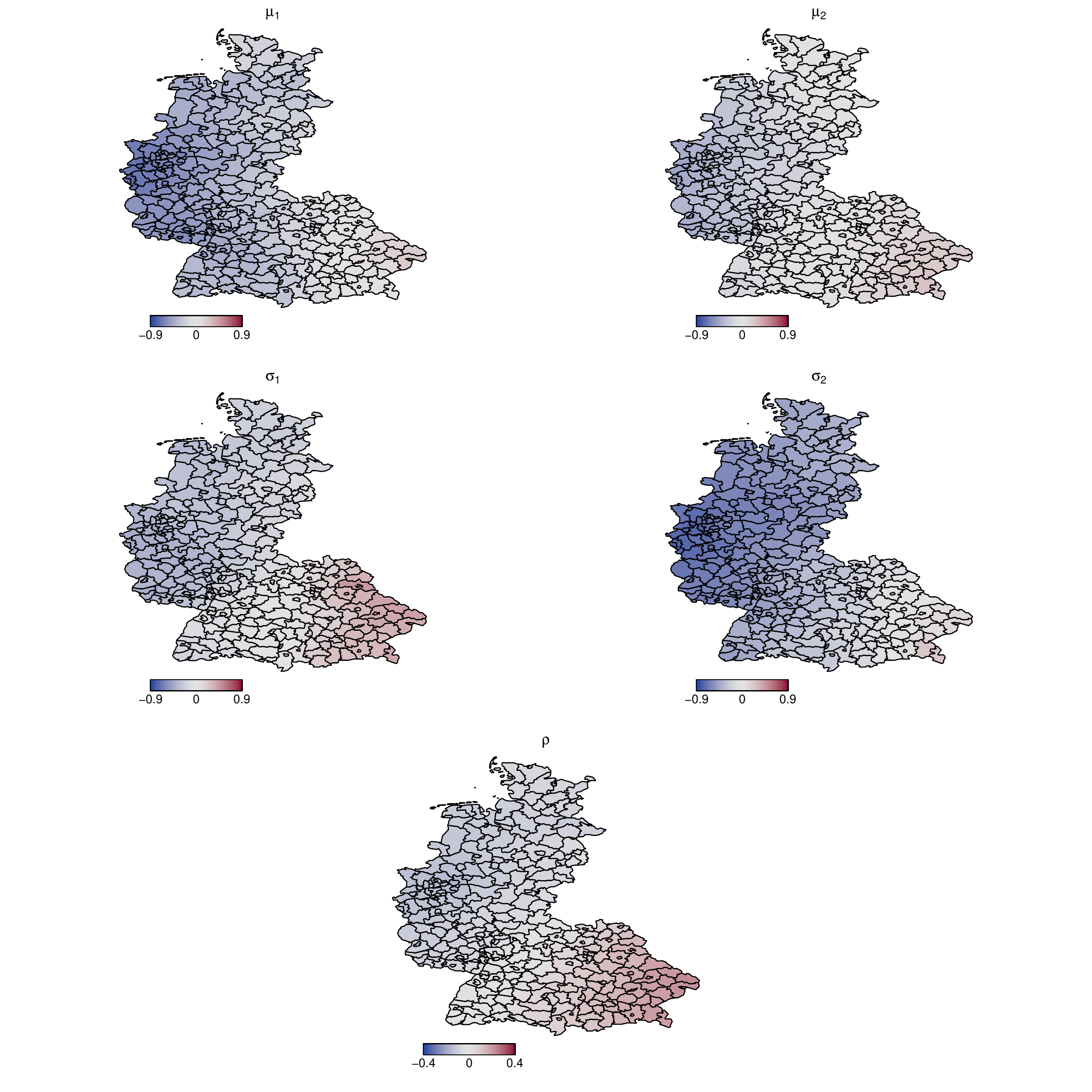} 
    \caption{The estimated spatial effects for one simulation run of the high-dimensional setting of the bivariate Gaussian regression model. }
\end{figure}

\clearpage
\newpage

\section{Biomedical applications}

\subsection{Genetic predisposition for chronic ischemic heart disease and high cholesterol}

\begin{figure}[h!]\centering
    \includegraphics[width = \textwidth]{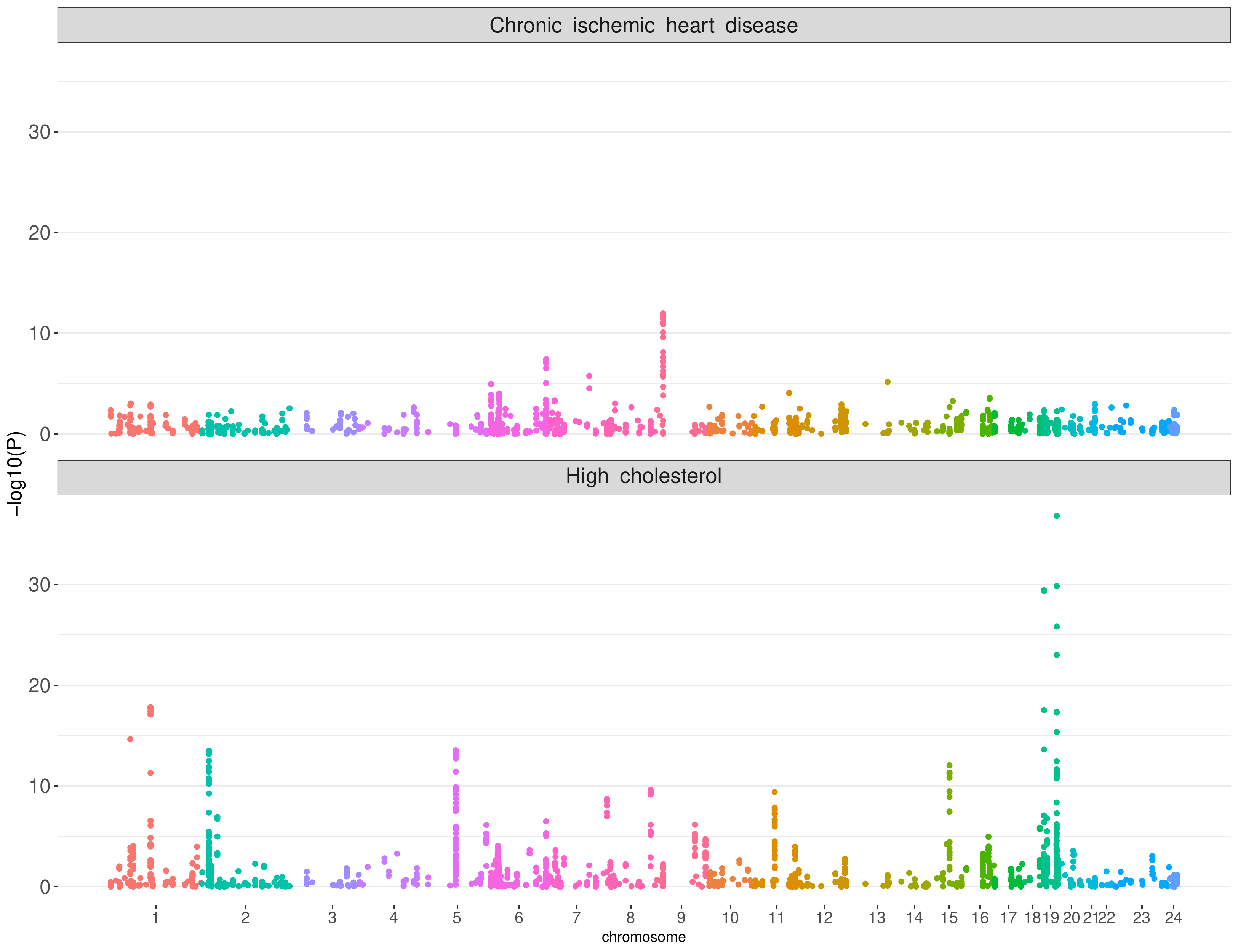} 
    \caption{Manhattan plot for the marginal association of the genetic variants on high cholesterol and chronic ischemic heart disease on the $-\log10(P)$ scale.}
\end{figure}

\newpage

\subsection{Risk factors for undernutrition in Nigeria}

\begin{table}[!ht]
    \centering
    \resizebox{\textwidth}{!}{%
    \begin{tabular}{l|clcl}
    \toprule
        Variable & Description &  \multicolumn{2}{l}{Type} & Mean (sd) \\ \midrule
        mage & mother's age &  \multicolumn{2}{l}{continuous} & 27.8 (6.8) \\ 
        mbmi & mother's bmi &  \multicolumn{2}{l}{continuous} & 23.1 (4.0) \\ 
        cage & child's age & \multicolumn{2}{l}{continuous} & 28.7 (17.2) \\ 
        edupartner & education years of mother's partner & \multicolumn{2}{l}{continuous} & 6.6 (5.8) \\ \midrule
        ~ & ~ & ~ & ~ & Percentage \\ \midrule
        bicycle & household has bicycle & binary & yes=1/no=0 & 21.7/78.3 \\ 
        car & household has car & binary & yes=1/no=0 & 9.5/90.5 \\ 
        cbirthorder1 & first child & binary & reference & 17.6/82.4 \\ 
        cbirthorder2 & second child & binary & yes=1/no=0 & 17.4/82.6 \\ 
        cbirthorder3 & third child & binary & yes=1/no=0 & 15.5/84.5 \\ 
        cbirthorder4 & fourth child & binary & yes=1/no=0 & 13.4/86.6 \\ 
        cbirthorder5 & fifth child & binary & yes=1/no=0 & 11.0/89.0 \\
        cbirthorder6 & sixth child & binary & yes=1/no=0 & 8.5/91.5 \\ 
        cbirthorder7 & seventh child & binary & yes=1/no=0 & 6.1/93.9 \\ 
        cbirthorder8 & eighth child & binary & yes=1/no=0 & 10.5/89.5 \\ 
        csex & child's sex & binary & female =1/male=0 & 49.6/50.4 \\ 
        ctwin & child a twin & binary & twin = 1/single birth = 0 & 2.8/97.2 \\ 
        electricity & household has electricity & binary & yes=1/no=0 & 48.3/51.7 \\ 
        motorcycle & household has motorcycle & binary & yes=1/no=0 & 41.2/58.8 \\ 
        mresidence & type of residence & binary & urban = 1/ rural = 0 & 65.4/34.6 \\ 
        munemployed & employment status of the mother & binary & employed= 1/unemployed = 0 & 72.4/27.6 \\ 
        radio & household has a radio & binary & yes=1/no=0 & 68.6/31.4 \\ 
        refrigerator & household has a refrigerator & binary & yes=1/no=0 & 16.7/83.3 \\ 
        television & household has a television & binary & yes=1/no=0 & 44.2/55.8 \\ \bottomrule
    \end{tabular}}
    \caption{Description of the explanatory variables for undernutrition in Nigeria.}
\end{table}

\begin{table}[h]
\centering
\begin{tabular}{lrrrrr}
  \toprule
  Covariates & $\mu_{\text{stunting}}$ & $\mu_{\text{wasting}}$ & $\sigma_{\text{stunting}}$ & $\sigma_{\text{wasting}}$ &  \multicolumn{1}{c}{$\rho$} \\  \midrule
  Intercept & -1.2667 & -0.7516 & 0.6301 & 0.2839 & -0.1728 \\ 
  bicycle & -0.0050 & -0.0185 & - & -0.0145 & 0.0319 \\ 
  car & 0.1379 & 0.0236 & - & -0.0374 & - \\ 
  cbirthorder2 & 0.0005 & 0.0427 & 0.0081 & 0.0380 & 0.0165 \\ 
  cbirthorder3 & -0.1090 & -0.0068 & 0.0089 & 0.0419 & 0.0034 \\ 
  cbirthorder4 & -0.1575 & -0.0497 & 0.0419 & - & 0.0095 \\ 
  cbirthorder5 & -0.1368 & -0.0521 & 0.0326 & 0.0019 & 0.0020 \\ 
  cbirthorder6 & -0.2185 & -0.0179 & 0.0363 & -0.0191 & 0.0269 \\ 
  cbirthorder7 & -0.2448 & -0.0319 & 0.0123 & -0.0158 & 0.0376 \\ 
  cbirthorder8 & -0.3185 & -0.0115 & 0.0255 & 0.0275 & 0.0282 \\ 
  csex & 0.1564 & 0.0388 & 0.0117 & 0.0008 & -0.0070 \\ 
  ctwin & -0.3672 & -0.1959 & 0.0436 & -0.0459 & - \\ 
  electricity & 0.0540 & - & 0.0050 & 0.0087 & - \\ 
  motorcycle & 0.0016 & 0.0070 & - & -0.0176 & -0.0104 \\ 
  mresidence & -0.0215 & 0.0624 & - & - & - \\ 
  munemployed & - & - & -0.0034 & -0.0340 & - \\ 
  radio & 0.0197 & - & -0.0060 & 0.0114 & - \\ 
  refrigerator & 0.1247 & - & - & - & 0.0763 \\ 
  television & 0.0820 & -0.0026 & -0.0401 & -0.0382 & 0.0016 \\ 
  \bottomrule
\end{tabular}
\caption{Results of the linear effects for \textit{stunting} and \textit{wasting} of the bivariate Gaussian regression model for the Nigeria data.}\label{Tab:Nigeria}
\end{table}

\end{document}